\documentclass{nature}[a4paper]

\usepackage{bm}
\usepackage{geometry}
\usepackage{authblk}
\usepackage{amsmath}
\usepackage{amssymb}
\usepackage{gensymb}
\usepackage{graphicx}
\usepackage{xcolor}
\usepackage[section]{placeins}
\makeatletter
\let\saved@includegraphics\includegraphics
\AtBeginDocument{\let\includegraphics\saved@includegraphics}
\renewenvironment*{figure}{\@float{figure}}{\end@float}
\makeatother
\usepackage[labelfont=bf]{caption}
\usepackage{cite}
\usepackage{soul}
\DeclareUnicodeCharacter{FF0C}{,}
\newcommand{\upcite}[1]{\textsuperscript{\scalebox{0.7}{\cite{#1}}}}

\title{\textcolor{blue}{Electric field switching of altermagnetic spin-splitting in multiferroic skyrmions} \footnotetext{$^\ast$  E-mail: chenxianzhe@fudan.edu.cn
\\}}

\author[1$\dagger$]{{Gui Wang}}
\author[2$\dagger$]{Yuhang Li}
\author[3$\dagger$]{\textcolor{blue}{Bin Li}}
\author[1$\ast$]{{Xianzhe Chen}}
\author[4]{Jianting Dong}
\author[5]{\textcolor{blue}{Weizhao Chen}}
\author[5]{\textcolor{blue}{Xiaobing Chen}}
\author[5]{\textcolor{blue}{Naifu Zheng}}
\author[6]{Maosen Guo}
\author[6]{Aomei Tong}
\author[7]{Hua Bai}
\author[8]{Hongrui Zhang}
\author[1]{Yifan Gao}
\author[1]{Kaiwen Shen}
\author[1]{Jiangyuan Zhu}
\author[9]{Jiahao Han}
\author[1]{Yingfen Wei}
\author[1]{Hao Jiang}
\author[1]{Xumeng Zhang}
\author[1]{Ming Wang}
\author[6]{Kebiao Xu}
\author[3]{\textcolor{blue}{Wu Shi}}
\author[10,14]{Pengfei Wang}
\author[4]{Jia Zhang}
\author[5,11]{\textcolor{blue}{Qihang Liu}}
\author[7]{Cheng Song}
\author[1]{Qi Liu}
\author[12,13,14]{Xincheng Xie}
\author[1]{Ming Liu}

\affil[1]{Frontier Institute of Chip and System, State Key Laboratory of Integrated Chips and Systems, Zhangjiang Fudan International Innovation Center, Fudan University, Shanghai 200433, China}
\affil[2]{School of Physics, Nankai University, Tianjin 300071, China}
\affil[3]{\textcolor{blue}{State Key Laboratory of Surface Physics and Institute for Nanoelectronic Devices and Quantum Computing, Fudan University, Shanghai 200433, China}}
\affil[4]{School of Physics and Wuhan National High Magnetic Field Center, Huazhong University of Science and Technology, Wuhan 430074, China}
\affil[5]{\textcolor{blue}{China Quantum Science Center of Guangdong-Hong Kong-Macao Greater Bay Area (Guangdong), Shenzhen 518045, China}
\affil[6]{CIQTEK Co,. Ltd. ，Hefei, 230026, China}}
\affil[7]{Key Laboratory of Advanced Materials (MOE), School of Materials Science and Engineering, Tsinghua University, Beijing 100084, China}
\affil[8]{Ningbo Institute of Materials Technology \& Engineering, Chinese Academy of Sciences, Ningbo 315201, China}
\affil[9]{Center for Science and Innovation in Spintronics, Tohoku University, 2-1-1 Katahira, 980-8577 Sendai, Japan}
\affil[10] {CAS Key Laboratory of Microscale Magnetic Resonance and School of Physical Sciences, University of Science and Technology of China, Hefei 230026, China}
\affil[11] {\textcolor{blue}{State Key laboratory of quantum functional materials, Department of Physics, and Guangdong Basic Research Center of Excellence for Quantum Science, Southern University of Science and Technology (SUSTech), Shenzhen 518055, China}}
\affil[12]{Interdisciplinary Center for Theoretical Physics and Information Sciences (ICTPIS), Fudan University, Shanghai 200433, China}
\affil[13]{International Center for Quantum Materials, Peking University, Beijing 100871, China}
\affil[14]{Hefei National Laboratory, Hefei 230088, China}

\graphicspath{{figures/}}

\date{}                   
\begin{document}
	
    \maketitle
	
	\vspace{10pt}
	
	\begin{abstract}
		\textbf{
        {\color{blue}Magnetic skyrmions are localized magnetic structures that retain their shape and stability over time, thanks to their topological nature\upcite{yu2010,jiang2015, jiang2017,woo2016,boulle2016,litzius2017,Tokura2021,Fert2017,liu2024,zhang2023}. Recent theoretical and experimental progress has laid the groundwork for understanding magnetic skyrmions characterized by negligible net magnetization and ultrafast dynamics\upcite{barker2016,zhang2016,gao2020,jani2021,legrand2020,amin2023,pham2024,Velez2022}.
        %{\color{red}Magnetic skyrmions are localized topological spin textures with remarkable stability and dynamic properties\upcite{yu2010,jiang2015,jiang2017,woo2016,boulle2016,litzius2017}, making them promising candidates for advanced spintronic applications. While traditional skyrmions have been extensively studied in materials with ferromagnetic or antiferromagnetic ordering\upcite{barker2016,zhang2016,gao2020,jani2021,legrand2020,pham2024},}
       Notably, skyrmions emerging in materials with altermagnetism\upcite{vsmejkal2020,vsmejkal2022,feng2022,krempasky2024,bai2022,chen2022,han2024,vsmejkal2023,Jungwirth2024}, a novel magnetic phase featuring lifted Kramers degeneracy—have remained unreported until now. In this study, we demonstrate that BiFeO$_3$, a multiferroic renowned for its strong coupling between ferroelectricity and magnetism\upcite{wang2003, Chu2008,Heron2014,gross2017}, can transit from a spin cycloid to a N\'eel-type skyrmion under antidamping spin-orbit torque at room temperature. Strikingly, the altermagnetic spin splitting within BiFeO$_3$ skyrmion can be reversed through the application of an electric field, revealed via the Circular photogalvanic effect. This quasiparticle, which possesses a neutral topological charge, holds substantial promise for diverse applications—most notably, enabling the development of unconventional computing systems with low power consumption and magnetoelectric controllability.}%such as the development of race-track memory\upcite{parkin2008} that minimizes unwanted stray fields.
        }\\
	\end{abstract}

    %Spin-splitting in electronic bands can be attributed to either relativistic effects, such as spin-orbit coupling (including Rashba and Dresselhaus types), or non-relativistic interactions like Zeeman splitting. 
    In conventional antiferromagnets\upcite{wadley2016,wadley2018,chen2018,zheng2022} with collinear moment alignments, the connection between magnetic sublattices relies on translation ($\mathcal{\tau}$) or parity ($\mathcal{P}$) operations, resulting in Kramers degeneracy, despite a disruption of pure time-reversal ($\mathcal{T}$) symmetry. In contrast, altermagnets connect their magnetic sublattices through rotation transformation, rather than $\mathcal{\tau}$ or $\mathcal{P}$ operations, which leads to a compensated alternating spin density in real space and simultaneously, nonrelativistic spin-momentum locking in reciprocal space\upcite{vsmejkal2022,libor2022Beyod,Smejkal2018}. %real-space symmetries, such as rotation or reflection, rather than time-reversal symmetries, which leads to a compensated alternating spin density in real space and simultaneously, spin-momentum locking in momentum space
     The unique alternating lifting of Kramers degeneracy in altermagnets gives rise to time-reversal symmetry-breaking magneto-responses\upcite{amin2024} %This unique alternating lifting of Kramers degeneracy in altermagnets gives rise to many interesting physical behaviors
     that are typically absent in conventional collinear antiferromagnets, e.g. anomalous Hall effect\upcite{vsmejkal2020,feng2022,han2024,reichlova2024}, tunneling magnetoresistance\upcite{chen2023,shao2021} and magnetic spin Hall effect\upcite{bai2022,bose2022}. In this article, we demonstrate the existence of altermagnetism in skyrmion phase of multiferroic BiFeO$_3$ (BFO). BFO is a model multiferroic that exhibits both antiferromagnetism (N\'eel temperature of $\sim 643$ K) and ferroelectricity (Curie temperature $\sim 1,100$ K and a large polarization $\sim 90$ $\mu C$ cm$^{-2}$ ) as well as intrinsic magnetoelectric coupling of these co-existing order parameters\upcite{bai2022,bose2022}. In the bulk, BFO possesses a rhombohedrally distorted perovskite structure in which a spin cycloid with a period length of $\sim 62$ nm is formed due to the intrinsic Dzyaloshinskii-Moriya interaction (DMI)\upcite{Ederer2005,Rahmedov2012}. In turn, BFO exhibits outstanding magnetoelectric properties that have made it the focus of numerous studies hoping to utilize it as a platform to achieve low-energy memory and logic devices\upcite{Chu2008,Heron2014,Huang2024,chai2024}. %{Since BiFeO$_3$ is a multiferroic consist of antiferromagnet and ferroelelctrics, the PT symmetry is therefore broken, which correspond to the nature of an altermagnet. However, due to the antiferromagnetic spin cycloid structure, the altermagnetic nature of BFO is eliminated due to the cannel of net Neel vector. Since the vector of The spin degeneracies can be lifted by the relativistic SOC in altermagnets with breaking the crystal inversion symmetry due to the ferroelectricity. Therefore, the basic state of BFO is an antiferromagnet state. However, if we break the spin cycloid state of BFO. The altermagnetic nature of BFO would evolve again. these spin degeneracies can be lifted by the relativistic SOC in altermagnets even without breaking the crystal inversion symmetry. Here, using NV diamond imaging, we directly identify the weak and strong altermagnetic LKSD in the band structure of BFO.}
    
\begin{figure}[t!]
    	\centering
    	\includegraphics[width=1\textwidth]{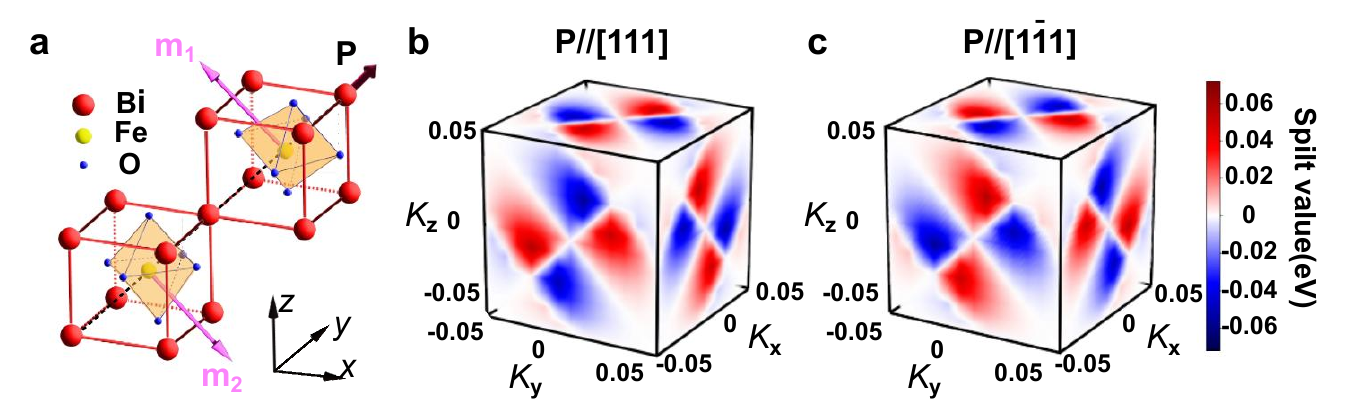}
    	\caption{\textbf{Concept of magnetoelectric altermagnet. %a, b,} Sketches of magnon propagation in BiFeO$_3$ lattice. The double well denotes the bi-stable states of ferroelectric polarizations. Upon a $180^\circ$ switching of the ferroelectric polarization from $[\bar{1} \bar{1} \bar{1}]$ in \textbf{a} to $[1 1 1]$ in \textbf{b} enabled by the negative charge center inversion, both the N\'{e}el vector and magnetization undergo a $180^\circ$ switching, leaving the N\'{e}el vector along the same axis.  (a) Crystal structure of BFO, with Bi atoms in red, Fe atoms in yellow, and O atoms in purple. The illustration shows the ferroelectric polarization along the [111] direction and the octahedral rotation of oxygen, which induces a Dzyaloshinskii-Moriya interaction (DMI) parallel to the polarization; the magnetic moments within the unit cell exhibit antiferromagnetic arrangement. (b) Spin-polarized band structure of BFO in K-space, revealing the K-space dependent spin splitting. (c) Density of states (DOS) plot for BFO.
        a, }{Crystal structure of BFO, with Bi atoms in red, Fe atoms in yellow, and O atoms in purple. The illustration shows the ferroelectric polarization $\mathbf{P}$ along the [111] direction and the octahedral rotation of oxygen, which induces a homogeneous DMI ($\mathbf{D}_1$) parallel to the polarization as well as an inhomogeneous DMI ($\mathbf{D}_2$); the magnetic moments within the unit cell exhibit antiferromagnetic arrangement. } \textbf{b, c,} {Spin-splitting energy dispersion of BFO. %Spin-polarized band structure of BFO. 
        {The splitting energy value is calculated for the highest band below the Fermi energy, revealing the spin-polarized band structure in $\mathbf{K}$-space}. The ferroelectric polarizations are along [111] (\textbf{b}) and [$1\bar{1}1$] (\textbf{c}) axis, respectively.}} 
    	\label{fig:1}
    	%\vspace{80pt}
    \end{figure}

    Figure~\ref{fig:1}a presents the crystal structure of BFO, highlighting the positions of the Bi, Fe, and O atoms, represented by red, yellow, and purple spheres, respectively. The ferroelectric polarization, denoted as $\mathbf{P}$, is aligned along the [111] direction. This polarization is accompanied by an octahedral rotation of the oxygen atoms, giving rise to an inhomogeneous DMI between the second nearest neighboring spins\upcite{Rahmedov2012} [$(\mathbf{D}_2\times\mathbf{e}_{ij})\cdot(\mathbf{S}_i\times\mathbf{S}_{j})$, where %$D_2$ denotes the DMI strength, $\mathbf{d}$ refers to the direction of the DMI that is aligned parallel to the polarization direction, and 
    $\mathbf{e}_{ij}$ is the vector connecting $\mathbf{S}_i$ and $\mathbf{S}_j$]. This DMI facilitates the formation of spin cycloid structure in BFO ground state and is crucial for the formation of the magnetic skyrmion. In addition, a homogeneous DMI [$\mathbf{D}_1\cdot(\mathbf{S}_i\times\mathbf{S}_{j}$)] also exists between nearest neighboring spins, which introduces a canting angle between neighboring spins and thus leads to a small net magnetization\upcite{Albrecht2010,Rahmedov2012}. 
    Since the combined parity and time-reversal ($\mathcal{PT}$) symmetry as well as spinor ($\mathcal{U}$) symmetry are explicitly breaking in BFO lattice due to the octahedral rotation of the oxygen atoms, BFO is anticipated to exhibit spin splitting in momentum space ($\mathbf{K}$-space). To reveal the altermagnetism of BFO, the spin-polarized energy band structures of BFO are calculated in $\mathbf{K}$-space without considering the spin-orbit coupling, where the ferroelectric polarization is directed along [111] (Fig.~\ref{fig:1}b). 
    %The spin-dependent splitting observed in the energy band structure demonstrates that the $\mathbf{K}$-space characteristics can be induced with the $\mathbf{PT}$-symmetry breaking inside BFO. 
    This splitting arises from the alternating crystal environment for the magnetic sublattice due to the octahedral rotation, leading to nonrelativistic momentum-dependent electronic states for spin-up and spin-down carriers%The splitting arises due to the interplay between spin-orbit coupling and the ferroelectric order, leading to momentum-dependent electronic states for spin-up and spin-down carriers {\color{red}{\color{red}The spin-dependent splitting observed in the energy band structure demonstrates the $\mathbf{K}$-space characteristics beyond the $\mathcal{PT}$ symmetry.}
    , and thus indicating that BFO is an altermagnet (Data Fig.~\ref{fig:Eig1}), {\color{blue}which is consistent with previous studies\upcite{bernardini2025,vsmejkal2024}}. 
    %{\color{red}This indicates that BFO is an altermagnet.}
    %{\color{blue} The argument on the altermagnetism of BFO is a little bit weak and contradictory to the definition of altermagnetism. A spin splitting of the energy bands is not necessarily indicative of altermagnetism because spin splitting is also allowed and observed in non-collinear AFM [Nature 626, 523 (2024), Phys. Rev. B 101, 220403(R) 2020]. Moreover, it seems like that altermagnetism only applies to collinear antiferromagnets, where the two spin sublattices after time reversal operation cannot be related by simple translation.} 
    The spin splitting of BFO is $\sim 0.1$ eV, in the same order as typical altermagnets e.g. Mn$_5$Si$_3$\upcite{han2024,krempasky2024}, MnF$_2$\upcite{Bhowal2024}. In contrast, Figure~\ref{fig:1}c shows the spin-splitting dispersion when the ferroelectric polarization points along the [$1\bar{1}1$] direction, where the polarity of spin-splitting flip sign. %, {\color{red}which reafirms the altermagnetic behavior of BFO}
    The reversed splitting features observed in this configuration highlight the potential for controlling altermagnetism through the ferroelectricity of BFO. %A schematic crystal structure of BiFeO3 is shown in Fig. 1c,d. The two crystal sublattices A and B of Fe atoms, whose magnetic moments order antiparallel below the transition temperature of 643 K, are connected by a non-symmorphic sixfold screw-axis rotation and are not connected by a translation or inversion. The resulting non-relativistic electronic structure of this altermagnet is of the g-wave type   with three spin-degenerate nodal planes parallel to the kz axis and crossing Γ and K points, and a fourth kz = 0 nodal plane (Fig. 1a).
    
\begin{figure}[ht!]
    	\centering
    	\includegraphics[width=1\textwidth]{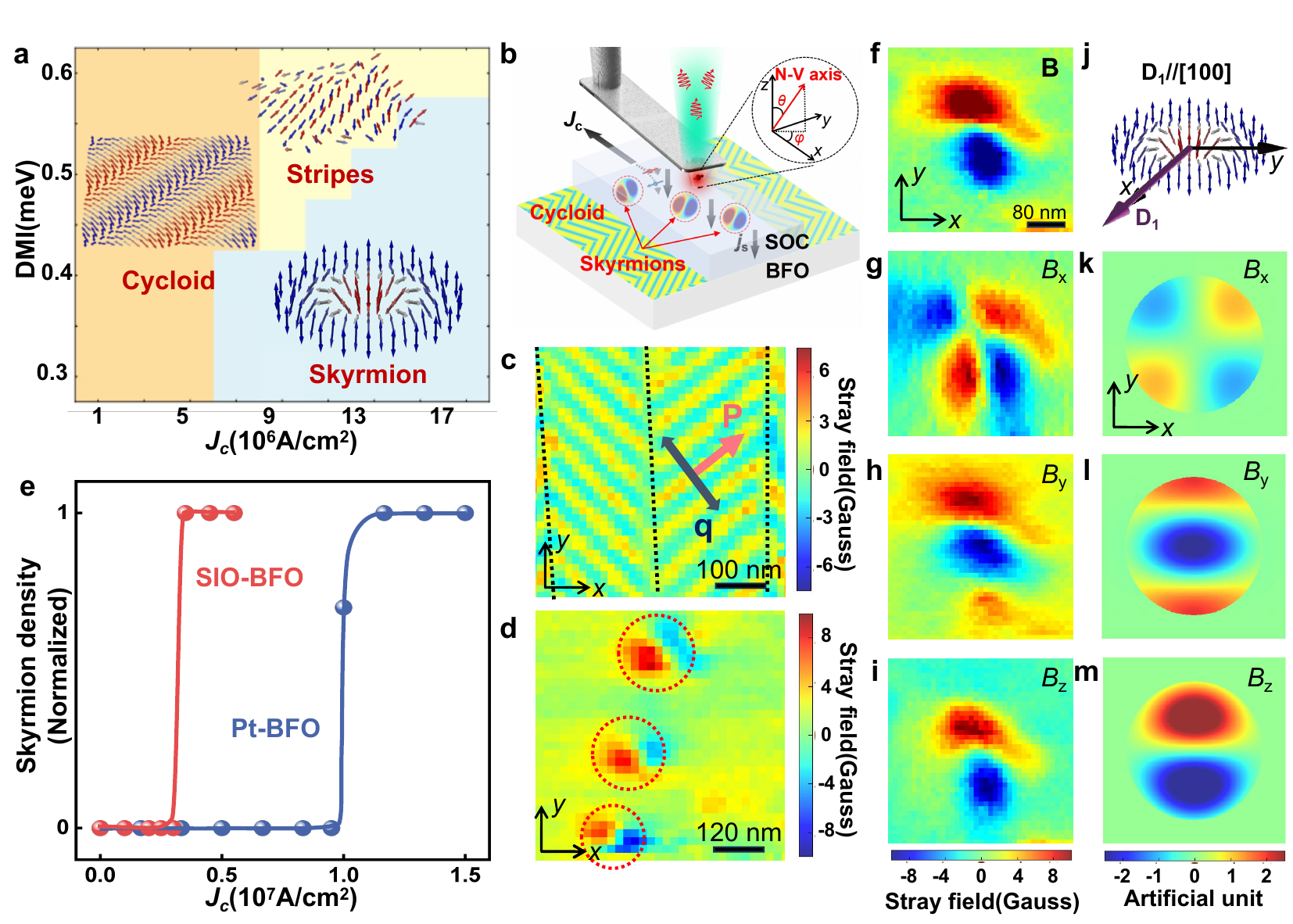}
    	\caption{\textbf{Magnetic structures and phase transition in BFO under spin current injection. a,} {\color{blue}Calculated phase diagram versus DMI and injected current densities $J_c$. Insets schematically illustrate the spin cycloid ground state (left), magnetic stripes (top right), and skyrmion (bottom right). \textbf{b,} Schematic of the experimental setup. The SOC layer is deposited on the BFO film and the scan nitrogen vacancy microscope (SNVM) is placed atop for imaging. A spin current $J_s$, generated via the spin Hall effect in the SOC layer, is injected into BFO, which triggers a phase transition from the cycloid state to the skyrmion. The directions of the SNVM axis and the sample structure (including the SOC layer and the BFO layer) are labeled in the figure. \textbf{c,} Magnetic domain image of the spin cycloid in BFO. Red and black arrows in (\textbf{c}) indicate the polarization direction $\mathbf{P}$ and the propagating direction $\mathbf{q}$, respectively, within a single domain. Black dashed lines indicate the domain boundaries. \textbf{d,} Image of skyrmions (highlighted in red circles) after injecting a current of $1.3\times 10^7$ A/cm$^2$ through the SOC layer. The scale bar is 100 nm in (\textbf{c}) while 120 nm in (\textbf{d}). \textbf{e,} Normalized skyrmion number per $\mu$m$^2$ versus applied current density $I$ for two different top electrodes SIO/BFO (red) and Pt/BFO (blue).} {\color{blue}\textbf{f,}} Normalized three-dimensional stray field map $\mathbf{B}(x,y)$ of a single skyrmion, obtained using SNVM on the sample surface, with $x$ and $y$ representing in-plane coordinates. In our experiment, the $x$-axis direction points along [100] while the  direction points along [010]. {\color{blue}\textbf{g-i,}} Reconstructed stray field components along the $x$-axis {\color{blue}(\textbf{g})}, $y$-axis {\color{blue}(\textbf{h})}, and $z$-axis {\color{blue}(\textbf{i})} directions based on panel {\color{blue}(\textbf{f})}. {\color{blue}\textbf{j,}} Simulated configuration of the skyrmion, with the homogeneous DMI ($\mathbf{D}_1$) direction aligned along the [100] direction. {\color{blue}\textbf{k-m,}} Simulated stray field distributions of the skyrmion in {\color{blue}(\textbf{f})} along the $x$-axis {\color{blue}(\textbf{k})}, $y$-axis {\color{blue}(\textbf{l})}, and $z$-axis {\color{blue}(\textbf{m})} directions. A one-to-one correspondence is observed between panels {\color{blue}(\textbf{g-i})} and {\color{blue}(\textbf{k-m})}, confirming the presence of skyrmions. All experiments are conducted at room temperature.
     }
    	\label{fig:2}
    	%\vspace{80pt}
    \end{figure}

    The magnetic ground state of BFO is a spin cycloid as illustrated in Fig.~\ref{fig:2}a (inset). 
    In this configuration, the antiferromagnetic moments rotate within a plane that is formed with ferroelectric polarization $\mathbf{P}$ and the propagation direction $\mathbf{q}$, which is also perpendicular to the direction of $\mathbf{D}_2$ {\color{blue}(Extended Data Fig.~\ref{fig:Eig2})}. However, the spin cycloid state is susceptible to external perturbations, particularly to the spin torque from spin currents\upcite{miron2011,liu2012}. When spin currents generated by the spin Hall effect\upcite{hoffmann2013} are injected into BFO, they interact with the existing magnetic structure\upcite{han2020,han2023,yoon2023}, leading to a transition from the cycloid state to a skyrmion state {\color{blue}(Extended Data Fig.~\ref{fig:Eig3}).} This transition is facilitated by the unique interplay between the spin currents and the DMI. The spin current introduces an anti-damping torque, which disturbs the spins and hence drives BFO into a high energy magnetic state. This state subsequently relaxes into the skyrmion configuration owing to the stabilizing influence of DMI\upcite{Li2023,Shashank2025Morphogenesis,Chaudron2024}. %which stabilizes the new skyrmion configuration. %{\color{blue}Can we elaborate a little bit more on the exact mechanism?}.
    {\color{blue}Figure~\ref{fig:2}a} presents a calculated phase diagram that delineates the boundaries among the cycloid, stripes and skyrmion states. The $x$-axis of the diagram represents the injected current density, while the $y$-axis indicates the strength of the inhomogeneous DMI ($\mathbf{D}_2$). The diagram clearly shows that a sufficient spin current can induce the transition from the spin cycloid to the skyrmion state. %Furthermore, the results indicate that higher DMI values result in lower critical spin current densities, suggesting that materials with strong DMI can facilitate easier transitions. 
    On the other hand, scanning nitrogen-vacancy microscopy (SNVM) {\color{blue}in full-B scanning mode is employed} to characterize the magnetic structure in BFO. {\color{blue}The experimental setup is illustrated in Fig.~\ref{fig:2}b, where a SOC layer utilized to generate spin current is deposited on the BFO film. The SNVM for imaging is further placed atop of the system.} Figure~\ref{fig:2}c features magnetic domain imaging of the spin cycloid in BFO, where the propagation direction $\mathbf{q}$ corresponds to the ferroelectric polarization $\mathbf{P}$ across different ferroelectric domains, showing strong magnetoelectric coupling in the BFO ground state {\color{blue}(Extended Data Fig.~\ref{fig:Eig4}).} In the same region, another magnetic domain image is taken following the injection of a current density of $1.3\times 10^7$ A/cm$^2$ in SOC (Pt) layer {\color{blue}as shown in Fig.~\ref{fig:2}d. In addition to coexisting non-topological magnetic texture displayed in the bottom, the topological spin textures with radius of $\sim 150\text{ nm}$ are clearly observed as highlighted by red circles. Figure~\ref{fig:2}e summarizes the observed topological states in this region as a function of applied electric current density. The critical switching current density for SrIrO$_3$ (SIO)-capped devices ($J_c$ = $3.5\times 10^6$ A/cm$^2$) is significantly lower than that of Pt-capped devices ($J_c$ = $1\times 10^7$ A/cm$^2$), demonstrating a 3-fold reduction that directly correlates with the giant spin Hall angle of SIO compared to Pt (Extended Data Fig.~\ref{fig:Eig5}). It should also be noted that the ferroelectric domains remain unchanged under current injection in SOC layer (Extended Data Fig.~\ref{fig:Eig6}).}
    %, and the topological spin textures are clearly observed as displayed in Fig.~\ref{fig:2}d. %{\color{blue}It might be worthwhile to have a methods section explaining how there experiments are performed.} 
    
    Next, we delve into the detailed analysis of the  skyrmion phase in BFO. This in principle can be revealed by the spatial distribution of the stray field uniquely bound to the magnetic spin texture. {\color{blue}In Fig.~\ref{fig:2}f,} we detect the three-dimensional stray field map $\mathbf{B}(x,y)$  of a single skyrmion obtained through scanning NV probe at the sample surface. The two-dot feature is strikingly distinct compared to the cycloid state. To elucidate the nature of this magnetic structure, we reconstruct the stray field components along the $x$, $y$ and $z$ directions, as shown in {\color{blue}Figs.~\ref{fig:2}(g-i).} In {\color{blue}Fig.~\ref{fig:2}g,} the $B_x$ component illustrates the spatial variation of the magnetic field along the horizontal axis, showcasing regions of both positive (red) and negative (blue) stray fields. Similarly, {\color{blue}Figure~\ref{fig:2}h} depicts the $B_y$ component, while {\color{blue}Fig.~\ref{fig:2}i} shows the $B_z$ component, which reflects the vertical behavior of the skyrmion’s stray field in relation to the sample surface. The distinct distributions of these components highlight the complexity of the stray field of this topological spin texture and its dependence on spatial orientation. In general, the stray field exhibits four bright spots in $B_x$ component, three bright spots in $B_y$ component while two bright spots in $B_z$ component. {\color{blue}Figures.~\ref{fig:2}(k-m)} display simulated stray field distributions of three corresponding components for a skyrmion with the $\mathbf{D}_1$ direction aligned along the [100] axis {\color{blue}(Fig.~\ref{fig:2}j)}. Notably, these simulated stray field distributions agree remarkably well with the experimental results shown in {\color{blue}Figs.~\ref{fig:2}(g-i),} which confirms the presence of skyrmions in our experiments. {\color{blue}Because the skyrmion size is smaller than the ferroelectric domain width, the zigzag pattern in Fig.2c does not interfere with the stablization of this altermagnetism. Moreover,} since the direction of the DMI can be modulated through the ferroelectric polarization, this can subsequently give rise to distinct magnetic configurations of skyrmions {\color{blue}(Extended Data Fig.~\ref{fig:Eig7}), which depends on $\mathbf{D}_1$ direction. }
    
\begin{figure}[ht!]
    	\centering
    	\includegraphics[width=1\textwidth]{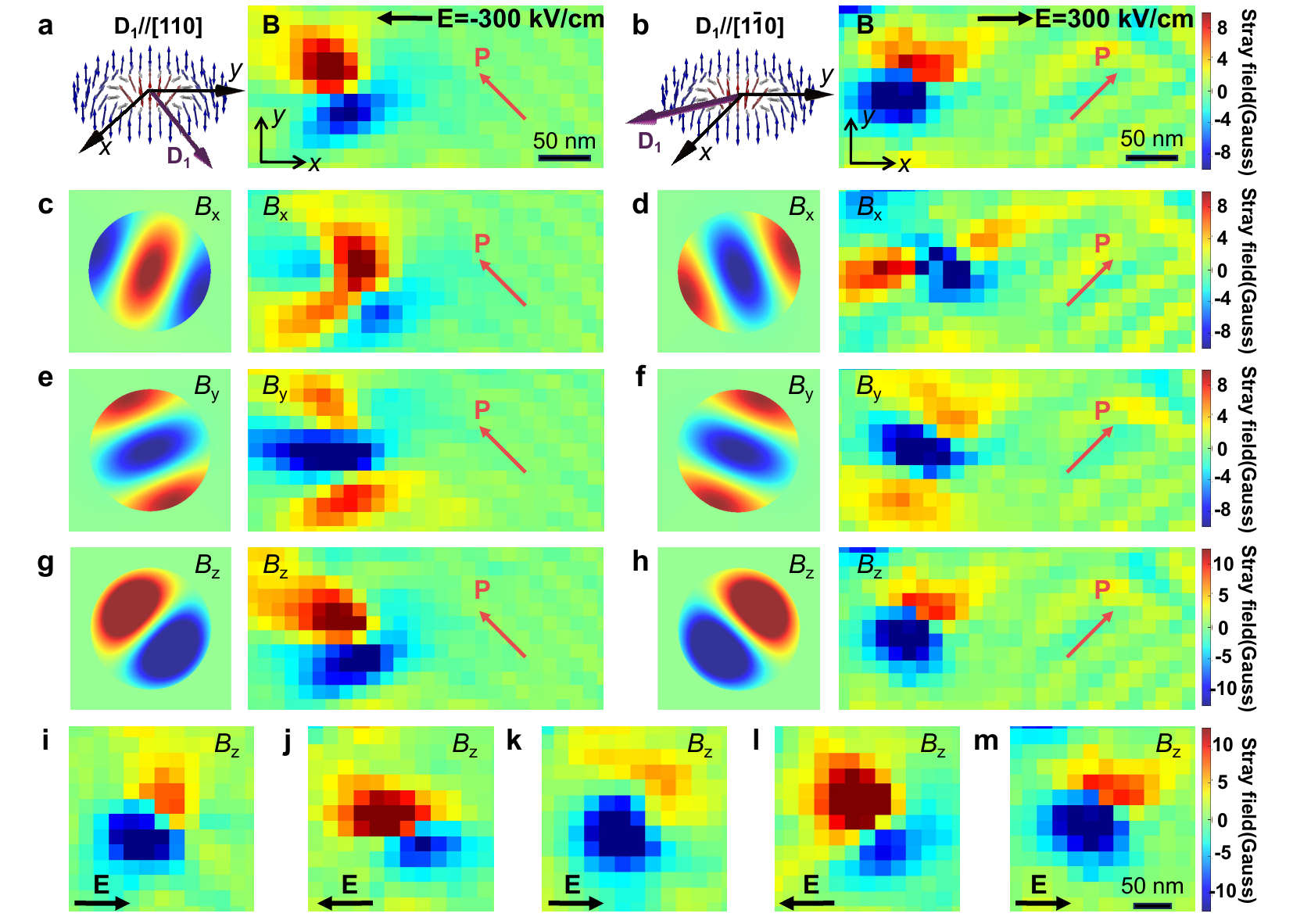}
    	\caption{\textcolor{blue}{{\textbf{Electric field manipulation of skyrmions. a,b } Left panels: Schematics of skyrmion with $\mathbf{D}_1$ along [$110$] (\textbf{a}) and [$1\bar{1}0$] (\textbf{b}) directions. Right panels: SNVM image of one skyrmion under E= -300 kV/cm (\textbf{a}) and E= 300 kV/cm (\textbf{b}), where corresponding polarization $\mathbf{P}$ aligns perpendicularly along [$110$] (\textbf{a}) and [$\bar{1}10$] (\textbf{b}). The comparison between simulated (left panels) and reconstructed stray field (right panels) along $x$-axis (\textbf{c,d}), $y$-axis (\textbf{e,f}), and $z$-axis (\textbf{g,h}) directions for corresponding skyrmions are shown below in (\textbf{c,e,g}) [based on panel (\textbf{a})] and (\textbf{d,f,h}) [based on panel (\textbf{b})]. Red and indigo arrows indicate the directions of $\mathbf{P}$ and $\mathbf{D}_1$, respectively. \textbf{i-m,} SNVM image represented by reconstructed stray field along $z$-axis after five consecutive electric field switching. The scale bar is 50 nm inlength.}}}
    	\label{fig:3}
    	%\vspace{80pt}
    \end{figure}

    {\color{blue} Distinct skyrmion configurations are stabilized by different ferroelectric polarization ($\mathbf{P}$) orientations in BFO. Given that $\mathbf{P}$ can be deterministically switched by an electric field, we further investigate direct electric-field control of the skyrmion phase, as demonstrated in Fig.~\ref{fig:3}. Experimental SNVM images capture the transformation of spin textures upon the electric field switching. Under a negative 300 kV/cm gate electric field, the local polarization rotates toward [$\bar{1}11$] axis, as evidenced by the spin cycloid phase, stabilizing a distinct skyrmion lattice configuration, where the direction of $\mathbf{D}_1$ is along [110] axis (Fig.~\ref{fig:3}a). When an equivalent positive electric field is applied, $\mathbf{P}$ switches to [111] axis, and the cycloid wave vector rotates 90° in-plane (Extended Data Fig.~\ref{fig:Eig8}). Simultaneously, the skyrmion pattern flips to the configuration with the DMI along the [$1\bar{1}0$] axis (Fig.~\ref{fig:3}b). This transformation indicates that the electric field can control the ferroelectric polarization in BFO, which in turns switches the skyrmion domains. The experimental observations are reproduced by our simulations with varying $\mathbf{D}_1$ (Figs.~\ref{fig:3}a-h). The simulated stray field maps (left panels in Figs.~\ref{fig:3}c-h) show near-perfect agreement with measured SNVM data (right panels in Figs.~\ref{fig:3}c-h). Next, we check the cyclability of the electric-field control of BFO skyrmion. SNVM images after successive $\mathbf{P}$ switches show reproducible flipping of the in-plane stray field components (Figs.~\ref{fig:3}i-m), while skyrmion size and stability remain unchanged. Notably, the entire process is driven solely by the electric field, without the assistance of an additional magnetic field or spin current. Fundamentally, the direct electric control of DMI orientation via polarization switching affirms an intrinsic magnetoelectric coupling. This reversible and non-volatile control of BFO skyrmion demonstrates a voltage-gated pathway to manipulate altermagnetic topological texture with ultralow power consumption.}
    
\begin{figure}[ht!]
    	\centering
    	\includegraphics[width=1\textwidth]{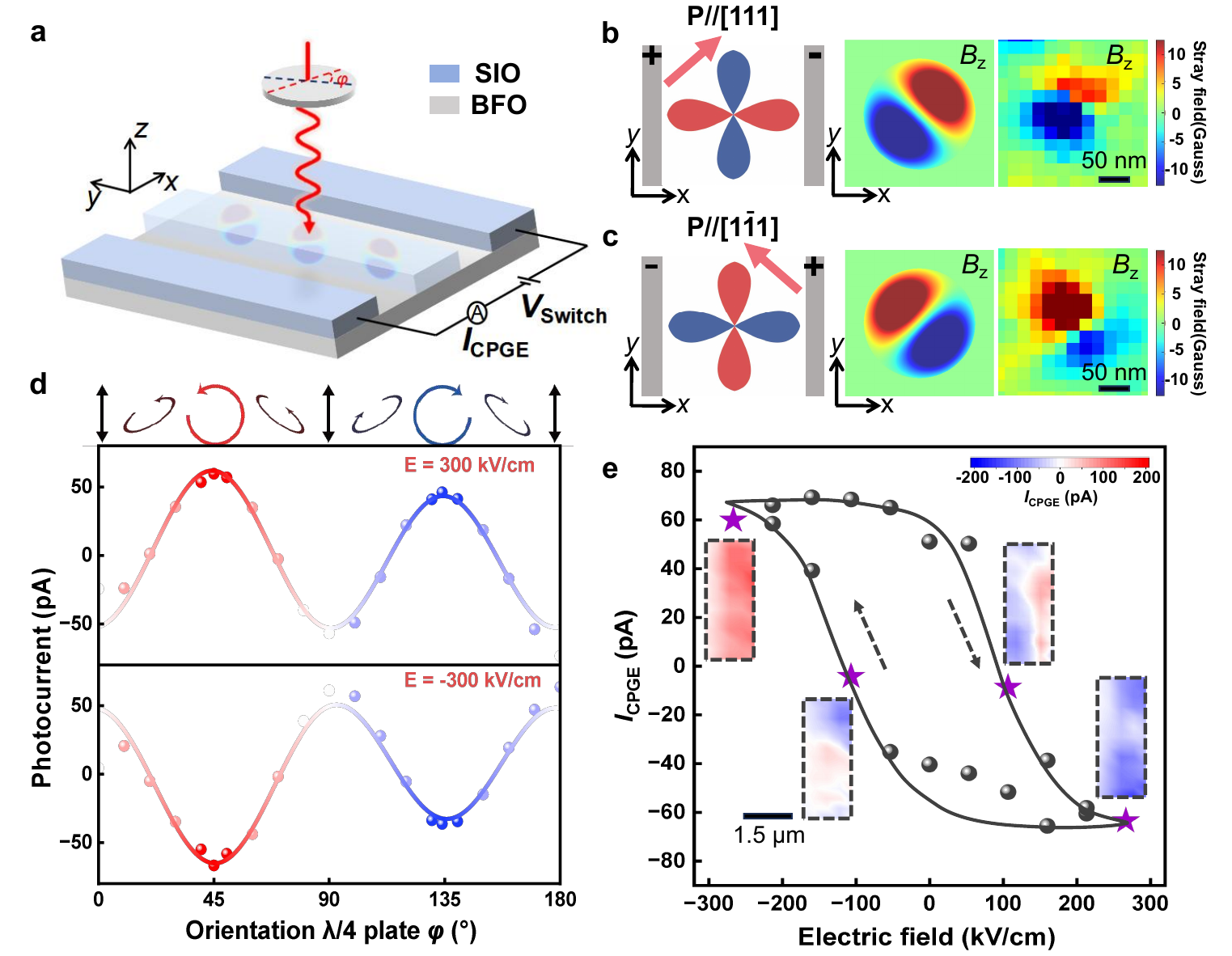}
    	\caption{\textcolor{blue}{\textbf{Electric-field switch of spin splitting, probed via the circular photogalvanic effect. a,} Schematics of photocurrent measurement. The spin photocurrent $I_{\mathrm{CPGE}}$ is generated from the circular photogalvanic effect under illumination of a 405 nm circularly polarized light. The voltage $V_{\mathrm{Switch}}$ is applied to induce an electric field for switching the ferroelectric polarization. \textbf{b, c,} Left panels schematically illustrate the two-dimensional projections of the electron energy bands on $k_x$-$k_y$ plane when the polarization direction is along [111] and [$1\bar{1}1$]. The electrodes are placed along the $y$-axis direction and are at a 45° angle to the polarization direction. Right panels in (\textbf{b}) and
         (\textbf{c}) show the skyrmion configurations after applying positive and negative electric fields to switch the polarization orientation of BFO. \textbf{d,} Photocurrent as a function of $\lambda/4$ plate orientation  measured with electrodes aligned along $y$-axis. Left and right -handed circular polarizations are represented as $\circlearrowleft$ and $\circlearrowright$. \textbf{e,} Responses of the net spin photocurrent ($I_{\mathrm{CPGE}}$) to electric-field modulation in skyrmion state within BFO. Insets display the mappings of $I_{\mathrm{CPGE}}$ under the corresponding electric field of the purple asterisks. 
    }}
    	\label{fig:4}
    	%\vspace{80pt}
    \end{figure}

    {\color{blue}The central advance lies in our ability to probe spin splitting through optical excitation and manipulate it with electric field. In contrast to spin cycloid, skyrmion exhibits non-vanish spin-splitting analogous to collinear state\upcite{Jin2024}. To access this spin-polarized band structure, we employ the circular photogalvanic effect (CPGE), a helicity-dependent optoelectronic response that maps spin-polarized carrier excitation to probe the spin splitting in BFO. As illustrated in Fig.~\ref{fig:4}a, circularly polarized light generates a photocurrent whose sign depends on the photon helicity. Light of a given helicity excites charges with spin up, while of opposite helicity excites charges with spin down\upcite{ganichev2003,mciver2012,duan2023,song2025}. The left panels in Figs.~\ref{fig:4}b and c schematically illustrate the two-dimensional projections of the electron energy bands on the $k_x-k_y$ plane when the polarization direction $\mathbf{P}$ is along [111] and $[11\overline{1}]$, respectively. The electrodes are placed along the $y$-axis direction and are at a $45\degree$ angle to the polarization direction. The two right panels in Figs. 4b and c show the different skyrmion configurations after applying positive and negative electric fields, where the skyrmion domains are controlled via the polarization switch of BFO. Crucially, we reveal the spin splitting of BFO skyrmion in Fig.~\ref{fig:4}d, which also link the CPGE response to the ferroelectric polarization state. A quarter-wave plate is used to change the light polarization. As shown in the top panel of Fig.~\ref{fig:4}d, under a positive gate electric field of 300 kV/cm, we observe a clear sinusoidal modulation of photocurrent as a function of the quarter-wave plate angle ($\varphi$). As $\varphi$ is varied, the light polarization gradually changes from linear to elliptical, and eventually to circular. The maximum current magnitude is recorded at $\varphi=45\degree$ for left-handed ($\circlearrowleft$) circular polarization and $\varphi=135\degree$ for right-handed ($\circlearrowright$) circular polarization. In the bottom panel of Fig.~\ref{fig:4}d, upon a negative electric field of identical magnitude, $\mathbf{P}$ is switched towards [$1\bar{1}1$]. The spin photocurrent polarity flips from positive to negative for the same photon helicity.}

    {\color{blue}To further reveal the spatially resolved spin-splitting in BFO, we map the photocurrent generated under $\circlearrowleft$ and $\circlearrowright$ circularly polarized light across the sample surface and extract the net spin photocurrent ($I_{\mathrm{CPGE}} = I_{\sigma}^{-}(\varphi=135^\circ) - I_{\sigma}^{+}(\varphi=45^\circ)$).
    % which serves as a direct evidence of momentum-space spin splitting arising from broken time-reversal and parity symmetries.
    As shown in Fig.~\ref{fig:4}e, $I_{\mathrm{CPGE}}$ exhibits a pronounced, reversible dependence on the applied electric fields, forming a well-defined electrical hysteresis loop at room temperature. The polarity of the $I_{\mathrm{CPGE}}$ switches sharply with ferroelectric polarization reversal, indicating tight coupling between the altermagnetic spin-splitting and the ferroelectric state. Insets display representative $I_{\mathrm{CPGE}}$ mappings after applying different electric fields (marked by purple asterisks), reflecting the optical spin current response behavior during the electric field switching process. We compare the spatially averaged $I_{\mathrm{CPGE}}$ in both the pristine spin cycloid\upcite{Knoche2021} and the induced skyrmion phases under alternatively applied positive/negative electric field (Extended Data Fig.~\ref{fig:Eig9}). The spin photocurrent from cycloid state exhibits a finite but unidirectional positive signal regardless of field polarity—consistent with the gyrotropy from crystal symmetry or canted moments\upcite{Knoche2021}, while the counterpart from skyrmion phase displays a robust antisymmetric dependence on the electric field direction. This polarity reversal originates from momentum-space spin splitting unique to real-space Berry curvature from skyrmion topology. This enables efficient helicity-dependent carrier excitation, whereas the cycloid's fully compensated N\'eel order suppresses such splitting. Thus, the observation of an electrically switchable, hysteretic spin photocurrent provides compelling evidence of the altermagnetic spin splitting within BFO skyrmions, which can also be reversed through the application of an electric field.
    }

    %The skyrmion Hall angle is calculated to be 29.8($\pm$7.5) degree, demonstrating the hallmark behavior of the skyrmion Hall effect and highlighting the unique topological properties of skyrmions within the altermagnet.

    {\color{blue}In summary, we have revealed the altermagnetism and its electric-field switch in the multiferroic BFO at room temperature, which is highly pursued theoretically\upcite{zhang2024,gu2025,duan2025,chen2025,song2022}. Under anti-damping spin-orbit torque, spin cycloid in BFO is transformed to skyrmion. BFO skyrmion exhibits non-vanishing altermagnetic spin splitting, as evidenced by our CPGE measurement, where the sign of the spin photocurrent can be reversed via electric field. These deepen our understanding of skyrmion behavior within altermagnets and showcase the potential of these topological structures for future applications in magnetic devices. For example, the electric-field switch of spin splitting in skyrmion has the potential to enable low-energy spintronic devices with minimal stray fields.}

    {\color{blue}\textit{Note added---}During our submission, we noted a relevant work\upcite{song2025}. Song {\textit{et.\ al.}}, broke new ground by demonstrating electrical switching of non-relativistic spin polarization in a $p$-wave magnet (NiI${}_2$) via the circular photogalvanic effect (CPGE).  NiI${}_2$ is a $p$-wave magnet, whose spin polarizations at $+k$ and $-k$ are opposite, connected by ${\mathcal{T}\tau}$ symmetry (time reversal followed by fractional translation). In contrast, for altermagnets, the spin polarization at the $+k$ and $-k$ states are the same, connected by $\mathcal{T}U$ symmetry (time reversal followed by spin reversal). We achieved electric-field switching of $g$-wave altermagnet BiFeO${}_3$ at room temperature. }
  
    \newpage

    %\FloatBarrier
    
    %\FloatBarrier
    
    %\FloatBarrier
    
    \FloatBarrier

    %\newpage
    \noindent\textbf{References}
    \bibliographystyle{naturemag}
    \bibliography{bibliography}

@Article{song2025,
author={Song, Qian
and Stavri{\'{c}}, Srdjan
and Barone, Paolo
and Droghetti, Andrea
and Antonenko, Daniil S.
and Venderbos, J{\"o}rn W. F.
and Occhialini, Connor A.
and Ilyas, Batyr
and Erge{\c{c}}en, Emre
and Gedik, Nuh
and Cheong, Sang-Wook
and Fernandes, Rafael M.
and Picozzi, Silvia
and Comin, Riccardo},
title={Electrical switching of a p-wave magnet},
journal={Nature},
year={2025},
month={Jun},
day={01},
volume={642},
number={8066},
pages={64-70},
issn={1476-4687},
doi={10.1038/s41586-025-09034-7}
}

@Article{song2022,
author={Song, Qian
and Occhialini, Connor A.
and Erge{\c{c}}en, Emre
and Ilyas, Batyr
and Amoroso, Danila
and Barone, Paolo
and Kapeghian, Jesse
and Watanabe, Kenji
and Taniguchi, Takashi
and Botana, Antia S.
and Picozzi, Silvia
and Gedik, Nuh
and Comin, Riccardo},
title={Evidence for a single-layer van der Waals multiferroic},
journal={Nature},
year={2022},
month={Feb},
day={01},
volume={602},
number={7898},
pages={601-605},
issn={1476-4687},
doi={10.1038/s41586-021-04337-x}
}

@article{duan2023,
  title={Berry curvature dipole generation and helicity-to-spin conversion at symmetry-mismatched heterointerfaces},
  author={Duan, Siyu and Qin, Feng and Chen, Peng and Yang, Xupeng and Qiu, Caiyu and Huang, Junwei and Liu, Gan and Li, Zeya and Bi, Xiangyu and Meng, Fanhao and others},
  journal={Nat. Nanotechnol.},
  volume={18},
  number={8},
  pages={867--874},
  year={2023},
  publisher={Nature Publishing Group UK London}
}

@article{mciver2012,
  title={Control over topological insulator photocurrents with light polarization},
  author={McIver, JW and Hsieh, David and Steinberg, Hadar and Jarillo-Herrero, Pablo and Gedik, Nuh},
  journal={Nat. Nanotechnol.},
  volume={7},
  number={2},
  pages={96--100},
  year={2012},
  publisher={Nature Publishing Group UK London}
}

@article{ganichev2003,
  title={Spin photocurrents in quantum wells},
  author={Ganichev, Sergey D and Prettl, Wilhelm},
  journal={Journal of physics: Condensed matter},
  volume={15},
  number={20},
  pages={R935},
  year={2003},
  publisher={IOP Publishing}
}

@article{zhang2024,
  title={Predictable gate-field control of spin in altermagnets with spin-layer coupling},
  author={Zhang, Run-Wu and Cui, Chaoxi and Li, Runze and Duan, Jingyi and Li, Lei and Yu, Zhi-Ming and Yao, Yugui},
  journal={Phys. Rev. Lett.},
  volume={133},
  number={5},
  pages={056401},
  year={2024},
  publisher={APS}
}

@article{chen2025,
  title={Electrical switching of altermagnetism},
  author={Chen, Yiyuan and Liu, Xiaoxiong and Lu, Hai-Zhou and Xie, XC},
  journal={Phys. Rev. Lett.},
  volume={135},
  number={1},
  pages={016701},
  year={2025},
  publisher={APS}
}

@article{knoche2021,
  title={{Anomalous circular bulk photovoltaic effect in BiFeO${}_3$ thin films with stripe-domain pattern}},
  author={Knoche, David S and Steimecke, Matthias and Yun, Yeseul and M{\"u}hlenbein, Lutz and Bhatnagar, Akash},
  journal={Nat. Commun.},
  volume={12},
  number={1},
  pages={282},
  year={2021},
  publisher={Nature Publishing Group UK London}
}

@article{gu2025,
  title={Ferroelectric switchable altermagnetism},
  author={Gu, Mingqiang and Liu, Yuntian and Zhu, Haiyuan and Yananose, Kunihiro and Chen, Xiaobing and Hu, Yongkang and Stroppa, Alessandro and Liu, Qihang},
  journal={Phys. Rev. Lett.},
  volume={134},
  number={10},
  pages={106802},
  year={2025},
  publisher={APS}
}

@article{duan2025,
  title={Antiferroelectric altermagnets: Antiferroelectricity alters magnets},
  author={Duan, Xunkai and Zhang, Jiayong and Zhu, Ziye and Liu, Yuntian and Zhang, Zhenyu and {\v{Z}}uti{\'c}, Igor and Zhou, Tong},
  journal={Phys. Rev. Lett.},
  volume={134},
  number={10},
  pages={106801},
  year={2025},
  publisher={APS}
}

@article{bernardini2025,
  title={Ruddlesden--Popper and perovskite phases as a material platform for altermagnetism},
  author={Bernardini, Fabio and Fiebig, Manfred and Cano, Andr{\'e}s},
  journal={J. Appl. Phys.},
  volume={137},
  number={10},
  year={2025},
  publisher={AIP Publishing}
}

@misc{vsmejkal2024,
      title={{Altermagnetic multiferroics and altermagnetoelectric effect}}, 
      author={{\v{S}}mejkal, Libor},
      year={2024},
      eprint={arXiv:2411.19928},
      archivePrefix={arXiv},
      primaryClass={cond-mat.mtrl-sci},
      url={https://arxiv.org/abs/2411.19928}, 
}

@article{yu2010,
  title={{Real-space observation of a two-dimensional skyrmion crystal}},
  author={Yu, XZ and Onose, Yoshinori and Kanazawa, Naoya and Park, Joung Hwan and Han, JH and Matsui, Yoshio and Nagaosa, Naoto and Tokura, Yoshinori},
  journal={Nature},
  volume={465},
  number={7300},
  pages={901--904},
  year={2010},
  publisher={Nature Publishing Group UK London}
}

@article{jiang2015,
  title={{Blowing magnetic skyrmion bubbles}},
  author={Jiang, Wanjun and Upadhyaya, Pramey and Zhang, Wei and Yu, Guoqiang and Jungfleisch, M Benjamin and Fradin, Frank Y and Pearson, John E and Tserkovnyak, Yaroslav and Wang, Kang L and Heinonen, Olle and others},
  journal={Science},
  volume={349},
  number={6245},
  pages={283--286},
  year={2015},
  publisher={American Association for the Advancement of Science}
}

@article{jiang2017,
  title={{Direct observation of the skyrmion Hall effect}},
  author={Jiang, Wanjun and Zhang, Xichao and Yu, Guoqiang and Zhang, Wei and Wang, Xiao and Benjamin Jungfleisch, M and Pearson, John E and Cheng, Xuemei and Heinonen, Olle and Wang, Kang L and others},
  journal={Nat. Phys.},
  volume={13},
  number={2},
  pages={162--169},
  year={2017},
  publisher={Nature Publishing Group UK London}
}

@article{woo2016,
  title={{Observation of room-temperature magnetic skyrmions and their current-driven dynamics in ultrathin metallic ferromagnets}},
  author={Woo, Seonghoon and Litzius, Kai and Kr{\"u}ger, Benjamin and Im, Mi-Young and Caretta, Lucas and Richter, Kornel and Mann, Maxwell and Krone, Andrea and Reeve, Robert M and Weigand, Markus and others},
  journal={Nat. Mater.},
  volume={15},
  number={5},
  pages={501--506},
  year={2016},
  publisher={Nature Publishing Group UK London}
}

@article{boulle2016,
  title={{Room-temperature chiral magnetic skyrmions in ultrathin magnetic nanostructures}},
  author={Boulle, Olivier and Vogel, Jan and Yang, Hongxin and Pizzini, Stefania and de Souza Chaves, Dayane and Locatelli, Andrea and Mente{\c{s}}, Tevfik Onur and Sala, Alessandro and Buda-Prejbeanu, Liliana D and Klein, Olivier and others},
  journal={Nat. Nanotechnol.},
  volume={11},
  number={5},
  pages={449--454},
  year={2016},
  publisher={Nature Publishing Group UK London}
}

@article{litzius2017,
  title={{Skyrmion Hall effect revealed by direct time-resolved X-ray microscopy}},
  author={Litzius, Kai and Lemesh, Ivan and Kr{\"u}ger, Benjamin and Bassirian, Pedram and Caretta, Lucas and Richter, Kornel and B{\"u}ttner, Felix and Sato, Koji and Tretiakov, Oleg A and F{\"o}rster, Johannes and others},
  journal={Nat. Phys.},
  volume={13},
  number={2},
  pages={170--175},
  year={2017},
  publisher={Nature Publishing Group UK London}
}

@article{barker2016,
  title={{Static and dynamical properties of antiferromagnetic skyrmions in the presence of applied current and temperature}},
  author={Barker, Joseph and Tretiakov, Oleg A},
  journal={Phys. Rev. Lett.},
  volume={116},
  number={14},
  pages={147203},
  year={2016},
  publisher={APS}
}

@article{zhang2016,
  title={{Antiferromagnetic skyrmion: stability, creation and manipulation}},
  author={Zhang, Xichao and Zhou, Yan and Ezawa, Motohiko},
  journal={Sci Rep},
  volume={6},
  number={1},
  pages={24795},
  year={2016},
  publisher={Nature Publishing Group UK London}
}

@article{gao2020,
  title={{Fractional antiferromagnetic skyrmion lattice induced by anisotropic couplings}},
  author={Gao, Shang and Rosales, H Diego and Gomez Albarracin, Flavia A and Tsurkan, Vladimir and Kaur, Guratinder and Fennell, Tom and Steffens, Paul and Boehm, Martin and {\v{C}}erm{\'a}k, Petr and Schneidewind, Astrid and others},
  journal={Nature},
  volume={586},
  number={7827},
  pages={37--41},
  year={2020},
  publisher={Nature Publishing Group UK London}
}

@article{jani2021,
  title={{Antiferromagnetic half-skyrmions and bimerons at room temperature}},
  author={Jani, Hariom and Lin, Jheng-Cyuan and Chen, Jiahao and Harrison, Jack and Maccherozzi, Francesco and Schad, Jonathon and Prakash, Saurav and Eom, Chang-Beom and Ariando, Ariando and Venkatesan, Thirumalai and others},
  journal={Nature},
  volume={590},
  number={7844},
  pages={74--79},
  year={2021},
  publisher={Nature Publishing Group UK London}
}

@article{legrand2020,
  title={{Room-temperature stabilization of antiferromagnetic skyrmions in synthetic antiferromagnets}},
  author={Legrand, William and Maccariello, Davide and Ajejas, Fernando and Collin, Sophie and Vecchiola, Aymeric and Bouzehouane, Karim and Reyren, Nicolas and Cros, Vincent and Fert, Albert},
  journal={Nat. Mater.},
  volume={19},
  number={1},
  pages={34--42},
  year={2020},
  publisher={Nature Publishing Group UK London}
}

@article{pham2024,
  title={{Fast current-induced skyrmion motion in synthetic antiferromagnets}},
  author={Pham, Van Tuong and Sisodia, Naveen and Di Manici, Ilaria and Urrestarazu-Larra{\~n}aga, Joseba and Bairagi, Kaushik and Pelloux-Prayer, Johan and Guedas, Rodrigo and Buda-Prejbeanu, Liliana D and Auffret, St{\'e}phane and Locatelli, Andrea and others},
  journal={Science},
  volume={384},
  number={6693},
  pages={307--312},
  year={2024},
  publisher={American Association for the Advancement of Science}
}

@article{vsmejkal2022,
  title={{Emerging research landscape of altermagnetism}},
  author={{\v{S}}mejkal, Libor and Sinova, Jairo and Jungwirth, Tomas},
  journal={Phys. Rev. X},
  volume={12},
  number={4},
  pages={040501},
  year={2022},
  publisher={APS}
}

@article{libor2022Beyod,
  title = {Beyond Conventional Ferromagnetism and Antiferromagnetism: A Phase with Nonrelativistic Spin and Crystal Rotation Symmetry},
  author = {\ifmmode \check{S}\else \v{S}\fi{}mejkal, Libor and Sinova, Jairo and Jungwirth, Tomas},
  journal = {Phys. Rev. X},
  volume = {12},
  issue = {3},
  pages = {031042},
  numpages = {16},
  year = {2022},
  month = {Sep},
  publisher = {American Physical Society},
  doi = {10.1103/PhysRevX.12.031042}
}

@Article{Smejkal2018,
author={{\v{S}}mejkal, Libor
and Mokrousov, Yuriy
and Yan, Binghai
and MacDonald, Allan H.},
title={Topological antiferromagnetic spintronics},
journal={Nat. Phys},
year={2018},
month={Mar},
day={01},
volume={14},
number={3},
pages={242-251},
issn={1745-2481},
doi={10.1038/s41567-018-0064-5}
}

@Article{Chaudron2024,
author={Chaudron, Arthur
and Li, Zixin
and Finco, Aurore
and Marton, Pavel
and Dufour, Pauline
and Abdelsamie, Amr
and Fischer, Johanna
and Collin, Sophie
and Dkhil, Brahim
and Hlinka, Jirka
and Jacques, Vincent
and Chauleau, Jean-Yves
and Viret, Michel
and Bouzehouane, Karim
and Fusil, St{\'e}phane
and Garcia, Vincent},
title={Electric-field-induced multiferroic topological solitons},
journal={Nat. Mater.},
year={2024},
month={Jul},
day={01},
volume={23},
number={7},
pages={905-911},
issn={1476-4660},
doi={10.1038/s41563-024-01890-4}
}

@article{
Shashank2025Morphogenesis,
author = {Shashank Kumar Ojha  and Pratap Pal  and Sergei Prokhorenko  and Sajid Husain  and Maya Ramesh  and Xinyan Li  and Deokyoung Kang  and Peter Meisenheimer  and Darrell G. Schlom  and Paul Stevenson  and Lucas Caretta  and Yousra Nahas  and Yimo Han  and Lane W. Martin  and Laurent Bellaiche  and Chang-Beom Eom  and Ramamoorthy Ramesh },
title = {Morphogenesis of spin cycloids in a noncollinear antiferromagnet},
journal = {Proc. Natl. Acad. Sci.},
volume = {122},
number = {17},
pages = {e2423298122},
year = {2025},
doi = {10.1073/pnas.2423298122},

}

@article{feng2022,
  title={{An anomalous Hall effect in altermagnetic ruthenium dioxide}},
  author={Feng, Zexin and Zhou, Xiaorong and {\v{S}}mejkal, Libor and Wu, Lei and Zhu, Zengwei and Guo, Huixin and Gonz{\'a}lez-Hern{\'a}ndez, Rafael and Wang, Xiaoning and Yan, Han and Qin, Peixin and others},
  journal={Nat. Electron.},
  volume={5},
  number={11},
  pages={735--743},
  year={2022},
  publisher={Nature Publishing Group UK London}
}

@article{krempasky2024,
  title={{Altermagnetic lifting of Kramers spin degeneracy}},
  author={Krempask{\`y}, Juraj and {\v{S}}mejkal, L and D’souza, SW and Hajlaoui, M and Springholz, G and Uhl{\'\i}{\v{r}}ov{\'a}, K and Alarab, F and Constantinou, PC and Strocov, V and Usanov, D and others},
  journal={Nature},
  volume={626},
  number={7999},
  pages={517--522},
  year={2024},
  publisher={Nature Publishing Group UK London}
}

@article{bai2022,
  title={{Observation of spin splitting torque in a collinear antiferromagnet RuO$_2$}},
  author={Bai, Hua and Han, Lei and Feng, XY and Zhou, YJ and Su, RX and Wang, Qian and Liao, LY and Zhu, WX and Chen, XZ and Pan, Feng and others},
  journal={Phys. Rev. Lett.},
  volume={128},
  number={19},
  pages={197202},
  year={2022},
  publisher={APS}
}

@article{Li2023,
  title = {{Multiferroic skyrmions in ${\mathrm{BiFeO}}_{3}$}},
  author = {Li, Z. and Chirac, T. and Tranchida, J. and Garcia, V. and Fusil, S. and Jacques, V. and Chauleau, J.-Y. and Viret, M.},
  journal = {Phys. Rev. Res.},
  volume = {5},
  issue = {4},
  pages = {043109},
  numpages = {6},
  year = {2023},
  month = {Nov},
  publisher = {American Physical Society},
  doi = {10.1103/PhysRevResearch.5.043109}
}

@article{han2024,
  title={{Electrical 180 switching of N{\'e}el vector in spin-splitting antiferromagnet}},
  author={Han, Lei and Fu, Xizhi and Peng, Rui and Cheng, Xingkai and Dai, Jiankun and Liu, Liangyang and Li, Yidian and Zhang, Yichi and Zhu, Wenxuan and Bai, Hua and others},
  journal={Sci. Adv.},
  volume={10},
  number={4},
  pages={eadn0479},
  year={2024},
  publisher={American Association for the Advancement of Science}
}

@article{vsmejkal2023,
  title={{Chiral magnons in altermagnetic RuO$_2$}},
  author={{\v{S}}mejkal, Libor and Marmodoro, Alberto and Ahn, Kyo-Hoon and Gonz{\'a}lez-Hern{\'a}ndez, Rafael and Turek, Ilja and Mankovsky, Sergiy and Ebert, Hubert and D’Souza, Sunil W and {\v{S}}ipr, Ond{\v{r}}ej and Sinova, Jairo and others},
  journal={Phys. Rev. Lett.},
  volume={131},
  number={25},
  pages={256703},
  year={2023},
  publisher={APS}
}

@article{Jin2024,
  title = {{Skyrmion Hall Effect in Altermagnets}},
  author = {Jin, Zhejunyu and Zeng, Zhaozhuo and Cao, Yunshan and Yan, Peng},
  journal = {Phys. Rev. Lett.},
  volume = {133},
  issue = {19},
  pages = {196701},
  numpages = {8},
  year = {2024},
  month = {Nov},
  publisher = {American Physical Society},
  doi = {10.1103/PhysRevLett.133.196701}
}

@Article{Dovzhenko2018,
author={Dovzhenko, Y.
and Casola, F.
and Schlotter, S.
and Zhou, T. X.
and B{\"u}ttner, F.
and Walsworth, R. L.
and Beach, G. S. D.
and Yacoby, A.},
title={{Magnetostatic twists in room-temperature skyrmions explored by nitrogen-vacancy center spin texture reconstruction}},
journal={Nat. Commun.},
year={2018},
month={Jul},
day={13},
volume={9},
number={1},
pages={2712},
issn={2041-1723},
doi={10.1038/s41467-018-05158-9}
}

@article{Rahmedov2012,
  title={{Magnetic Cycloid of {B}i{F}e{O}$_3$ from Atomistic Simulations}},
  author={Rahmedov, Dovran and Wang, Dawei and Íñiguez, Jorge and Bellaiche, Laurent},
  journal={Phys. Rev. Lett.},
  volume={109},
  pages={037207},
  year={2012},
  publisher={American Physical Society}
}

@article{huang2024,
  title={{Manipulating chiral spin transport with ferroelectric polarization}},
  author={Huang, Xiaoxi and Chen, Xianzhe and Li, Yuhang and Mangeri, John and Zhang, Hongrui and Ramesh, Maya and Taghinejad, Hossein and Meisenheimer, Peter and Caretta, Lucas and Susarla, Sandhya and others},
  journal={Nat. Mater.},
  pages={1--7},
  year={2024},
  publisher={Nature Publishing Group UK London}
}

@article{chen2018,
  title={{Antidamping-torque-induced switching in biaxial antiferromagnetic insulators}},
  author={Chen, X Z and Zarzuela, R and Zhang, J and Song, C and Zhou, X F and Shi, G Y and Li, F and Zhou, H A and Jiang, W J and Pan, F and others},
  journal={Phys. Rev. Lett.},
  volume={120},
  number={20},
  pages={207204},
  year={2018},
  publisher={APS}
}

@article{reichlova2024,
  title={{Observation of a spontaneous anomalous Hall response in the Mn$_5$Si$_3$ d-wave altermagnet candidate}},
  author={Reichlova, Helena and Lopes Seeger, Rafael and Gonz{\'a}lez-Hern{\'a}ndez, Rafael and Kounta, Ismaila and Schlitz, Richard and Kriegner, Dominik and Ritzinger, Philipp and Lammel, Michaela and Leivisk{\"a}, Miina and Birk Hellenes, Anna and others},
  journal={Nat. Commun.},
  volume={15},
  number={1},
  pages={4961},
  year={2024},
  publisher={Nature Publishing Group UK London}
}

@article{bose2022,
  title={{Tilted spin current generated by the collinear antiferromagnet ruthenium dioxide}},
  author={Bose, Arnab and Schreiber, Nathaniel J and Jain, Rakshit and Shao, Ding-Fu and Nair, Hari P and Sun, Jiaxin and Zhang, Xiyue S and Muller, David A and Tsymbal, Evgeny Y and Schlom, Darrell G and others},
  journal={Nat. Electron.},
  volume={5},
  number={5},
  pages={267--274},
  year={2022},
  publisher={Nature Publishing Group UK London}
}

@article{shao2021,
  title={{Spin-neutral currents for spintronics}},
  author={Shao, Ding-Fu and Zhang, Shu-Hui and Li, Ming and Eom, Chang-Beom and Tsymbal, Evgeny Y},
  journal={Nat. Commun.},
  volume={12},
  number={1},
  pages={7061},
  year={2021},
  publisher={Nature Publishing Group UK London}
}

@article{chen2022,
  title={{Control of spin current and antiferromagnetic moments via topological surface state}},
  author={Chen, Xianzhe and Bai, Hua and Ji, Yuchen and Zhou, Yongjian and Liao, Liyang and You, Yunfeng and Zhu, Wenxuan and Wang, Qian and Han, Lei and Liu, Xiaoyang and others},
  journal={Nat. Electron.},
  volume={5},
  number={9},
  pages={574--578},
  year={2022},
  publisher={Nature Publishing Group UK London}
}

@article{chen2023,
  title={{Octupole-driven magnetoresistance in an antiferromagnetic tunnel junction}},
  author={Chen, Xianzhe and Higo, Tomoya and Tanaka, Katsuhiro and Nomoto, Takuya and Tsai, Hanshen and Idzuchi, Hiroshi and Shiga, Masanobu and Sakamoto, Shoya and Ando, Ryoya and Kosaki, Hidetoshi and others},
  journal={Nature},
  volume={613},
  number={7944},
  pages={490--495},
  year={2023},
  publisher={Nature Publishing Group UK London}
}

@article{Ederer2005,
  title={{Weak ferromagnetism and magnetoelectric coupling in bismuth ferrite}},
  author={Ederer, Claude and Spaldin, Nicola A.},
  journal={Phys. Rev. B},
  volume={71},
  number={6},
  pages={060401(R)},
  year={2005},
  publisher={APS}
}

@article{liu2012,
  title={{Spin-torque switching with the giant spin {Hall} effect of tantalum}},
  author={Liu, Luqiao and Pai, Chi-Feng and Li, Y and Tseng, HW and Ralph, DC and Buhrman, RA},
  journal={Science},
  volume={336},
  number={6081},
  pages={555--558},
  year={2012},
  publisher={American Association for the Advancement of Science}
}

@article{wang2003,
  title={{Epitaxial {B}i{F}e{O}$_3$ multiferroic thin film heterostructures}},
  author={Wang, JBNJ and Neaton, JB and Zheng, H and Nagarajan, V and Ogale, SB and Liu, B and Viehland, D and Vaithyanathan, V and Schlom, DG and Waghmare, UV and others},
  journal={Science},
  volume={299},
  number={5613},
  pages={1719--1722},
  year={2003},
  publisher={American Association for the Advancement of Science}
}

@article{Chu2008,
  title={{Electric-field control of local ferromagnetism using a magnetoelectric multiferroic}},
  author={Chu, Ying-Hao and Martin, Lane W. and Holcomb, Mikel B. and Gajek, Martin and Han, Shu-Jen and He, Qing and Balke, Nina and Yang, Chan-Ho and Lee, Don-Koun and Hu, Wei and Zhan, Qian and Yang, Pei-Ling and Rodriguez, Arantxa Frail and Scholl, Andreas and Wang, Shan X and Ramesh. R},
  journal={Nature Mater.},
  volume={7},
  pages={478},
  year={2008},
  publisher={Nature Publishing Group}}

@article{1,
  title={{Temperature dependence of the crystal and magnetic structure of {BiFeO$_3$}}},
  author={Fischer, P and Polomska, M. and Sosnowska, I and Szymanski, M},
  journal={J. Phys. C: Solid St. Phys.},
  volume={19},
  pages={1931},
  year={1980},
  publisher={IOP Science}
  }

@article{Heron2014,
  title={{Deterministic switching of ferromagnetism at room temperature using an electric field}},
  author={Heron, JT and Bosse, JL and He, Q and Gao, Y and Trassin, Morgan and Ye, L and Clarkson, JD and Wang, C and Liu, Jian and Salahuddin, S and others},
  journal={Nature},
  volume={516},
  number={7531},
  pages={370--373},
  year={2014},
  publisher={Nature Publishing Group}
}

@article{gross2017,
  title={{Real-space imaging of non-collinear antiferromagnetic order with a single-spin magnetometer}},
  author={Gross, Isabell and Akhtar, W and Garcia, V and Mart{\'\i}nez, L J and Chouaieb, Saddem and Garcia, K and Carr{\'e}t{\'e}ro, C and Barth{\'e}l{\'e}my, A and Appel, P and Maletinsky, P and others},
  journal={Nature},
  volume={549},
  number={7671},
  pages={252--256},
  year={2017},
  publisher={Nature Publishing Group}
}

@Article{Tan2024,
author={Tan, Anthony K. C.
and Jani, Hariom
and H{\"o}gen, Michael
and Stefan, Lucio
and Castelnovo, Claudio
and Braund, Daniel
and Geim, Alexandra
and Mechnich, Annika
and Feuer, Matthew S. G.
and Knowles, Helena S.
and Ariando, Ariando
and Radaelli, Paolo G.
and Atat{\"u}re, Mete},
title={{Revealing emergent magnetic charge in an antiferromagnet with diamond quantum magnetometry}},
journal={Nat. Mater.},
year={2024},
month={Feb},
day={01},
volume={23},
number={2},
pages={205-211},
issn={1476-4660},
doi={10.1038/s41563-023-01737-4}
}

@Article{Tokura2021,
author={Tokura, Yoshinori
and Kanazawa, Naoya},
title={{Magnetic Skyrmion Materials}},
journal={Chemical Reviews},
year={2021},
month={Mar},
day={10},
publisher={American Chemical Society},
volume={121},
number={5},
pages={2857-2897},
issn={0009-2665},
doi={10.1021/acs.chemrev.0c00297}
}

@article{Bhowal2024,
  title = {{Ferroically Ordered Magnetic Octupoles in $d$-Wave Altermagnets}},
  author = {Bhowal, Sayantika and Spaldin, Nicola A.},
  journal = {Phys. Rev. X},
  volume = {14},
  issue = {1},
  pages = {011019},
  numpages = {19},
  year = {2024},
  month = {Feb},
  publisher = {American Physical Society},
  doi = {10.1103/PhysRevX.14.011019}
}

@Article{Fert2017,
author={Fert, Albert
and Reyren, Nicolas
and Cros, Vincent},
title={Magnetic skyrmions: advances in physics and potential applications},
journal={Nat. Rev. Mater.},
year={2017},
month={Jun},
day={13},
volume={2},
number={7},
pages={17031},
issn={2058-8437},
doi={10.1038/natrevmats.2017.31}
}

@misc{Jungwirth2024,
      title={{Altermagnets and beyond: Nodal magnetically-ordered phases}}, 
      author={Tomas Jungwirth and Rafael M. Fernandes and Jairo Sinova and Libor Smejkal},
      year={2024},
      eprint={arXiv:2409.10034},
      archivePrefix={arXiv},
      primaryClass={cond-mat.mtrl-sci},
      url={https://arxiv.org/abs/2409.10034}, 
}

@article{Albrecht2010,
  title = {{Ferromagnetism in multiferroic ${\text{BiFeO}}_{3}$ films: A first-principles-based study}},
  author = {Albrecht, D. and Lisenkov, S. and Ren, Wei and Rahmedov, D. and Kornev, Igor A. and Bellaiche, L.},
  journal = {Phys. Rev. B},
  volume = {81},
  issue = {14},
  pages = {140401},
  numpages = {4},
  year = {2010},
  month = {Apr},
  publisher = {American Physical Society},
  doi = {10.1103/PhysRevB.81.140401}
}

@article{chai2024,
  title={Voltage control of multiferroic magnon torque for reconfigurable logic-in-memory},
  author={Chai, Yahong and Liang, Yuhan and Xiao, Cancheng and Wang, Yue and Li, Bo and Jiang, Dingsong and Pal, Pratap and Tang, Yongjian and Chen, Hetian and Zhang, Yuejie and others},
  journal={Nat. Commun.},
  volume={15},
  number={1},
  pages={5975},
  year={2024},
  publisher={Nature Publishing Group UK London}
}

@article{wadley2016,
  title={Electrical switching of an antiferromagnet},
  author={Wadley, Peter and Howells, Bryn and {\v{Z}}elezn{\`y}, J and Andrews, Carl and Hills, Victoria and Campion, Richard P and Nov{\'a}k, Vit and Olejn{\'\i}k, K and Maccherozzi, F and Dhesi, SS and others},
  journal={Science},
  volume={351},
  number={6273},
  pages={587--590},
  year={2016},
  publisher={American Association for the Advancement of Science}
}

@article{wadley2018,
  title={Current polarity-dependent manipulation of antiferromagnetic domains},
  author={Wadley, Peter and Reimers, Sonka and Grzybowski, Michal J and Andrews, Carl and Wang, Mu and Chauhan, Jasbinder S and Gallagher, Bryan L and Campion, Richard P and Edmonds, Kevin W and Dhesi, Sarnjeet S and others},
  journal={Nat. Nanotechnol.},
  volume={13},
  number={5},
  pages={362--365},
  year={2018},
  publisher={Nature Publishing Group UK London}
}

@article{amin2023,
  title={Antiferromagnetic half-skyrmions electrically generated and controlled at room temperature},
  author={Amin, OJ and Poole, SF and Reimers, S and Barton, LX and Dal Din, A and Maccherozzi, F and Dhesi, SS and Nov{\'a}k, V and Krizek, F and Chauhan, JS and others},
  journal={Nat. Nanotechnol.},
  volume={18},
  number={8},
  pages={849--853},
  year={2023},
  publisher={Nature Publishing Group UK London}
}

@Article{amin2024,
author={Amin, O. J.
and Dal Din, A.
and Golias, E.
and Niu, Y.
and Zakharov, A.
and Fromage, S. C.
and Fields, C. J. B.
and Heywood, S. L.
and Cousins, R. B.
and Maccherozzi, F.
and Krempask{\'y}, J.
and Dil, J. H.
and Kriegner, D.
and Kiraly, B.
and Campion, R. P.
and Rushforth, A. W.
and Edmonds, K. W.
and Dhesi, S. S.
and {\v{S}}mejkal, L.
and Jungwirth, T.
and Wadley, P.},
title={Nanoscale imaging and control of altermagnetism in MnTe},
journal={Nature},
year={2024},
month={Dec},
day={01},
volume={636},
number={8042},
pages={348-353},
issn={1476-4687},
doi={10.1038/s41586-024-08234-x}
}

@article{vsmejkal2020,
  title={{Crystal time-reversal symmetry breaking and spontaneous Hall effect in collinear antiferromagnets}},
  author={{\v{S}}mejkal, Libor and Gonz{\'a}lez-Hern{\'a}ndez, Rafael and Jungwirth, Tom{\'a}{\v{s}} and Sinova, Jairo},
  journal={Sci. Adv.},
  volume={6},
  number={23},
  pages={eaaz8809},
  year={2020},
  publisher={American Association for the Advancement of Science}
}

@article{hoffmann2013,
  title={{Spin Hall effects in metals}},
  author={Hoffmann, Axel},
  journal={IEEE Trans. Magn.},
  volume={49},
  number={10},
  pages={5172--5193},
  year={2013},
  publisher={IEEE}
}

@article{liu2024,
  title={Magnetic Skyrmions above Room Temperature in a van der Waals Ferromagnet {Fe${}_3$GaTe${}_2$}},
  author={Liu, Chen and Zhang, Senfu and Hao, Hongyuan and Algaidi, Hanin and Ma, Yinchang and Zhang, Xi-Xiang},
  journal={Adv. Mater.},
  volume={36},
  number={18},
  pages={2311022},
  year={2024},
  publisher={Wiley Online Library}
}

@article{zheng2022,
  title={High-Efficiency Magnon-Mediated Magnetization Switching in All-Oxide Heterostructures with Perpendicular Magnetic Anisotropy},
  author={Zheng, Dongxing and Lan, Jin and Fang, Bin and Li, Yan and Liu, Chen and Ledesma-Martin, J Omar and Wen, Yan and Li, Peng and Zhang, Chenhui and Ma, Yinchang and others},
  journal={Adv. Mater.},
  volume={34},
  number={34},
  pages={2203038},
  year={2022},
  publisher={Wiley Online Library}
}

@article{zhang2023,
  title={Room-Temperature Magnetic Skyrmions and Large Topological Hall Effect in Chromium Telluride Engineered by Self-Intercalation},
  author={Zhang, Chenhui and Liu, Chen and Zhang, Junwei and Yuan, Youyou and Wen, Yan and Li, Yan and Zheng, Dongxing and Zhang, Qiang and Hou, Zhipeng and Yin, Gen and others},
  journal={Adv. Mater.},
  volume={35},
  number={1},
  pages={2205967},
  year={2023},
  publisher={Wiley Online Library}
}

@article{han2020,
  title={Birefringence-like spin transport via linearly polarized antiferromagnetic magnons},
  author={Han, Jiahao and Zhang, Pengxiang and Bi, Zhen and Fan, Yabin and Safi, Taqiyyah S and Xiang, Junxiang and Finley, Joseph and Fu, Liang and Cheng, Ran and Liu, Luqiao},
  journal={Nat. Nanotechnol.},
  volume={15},
  number={7},
  pages={563--568},
  year={2020},
  publisher={Nature Publishing Group UK London}
}

@article{han2023,
  title={Coherent antiferromagnetic spintronics},
  author={Han, Jiahao and Cheng, Ran and Liu, Luqiao and Ohno, Hideo and Fukami, Shunsuke},
  journal={Nat. Mater.},
  volume={22},
  number={6},
  pages={684--695},
  year={2023},
  publisher={Nature Publishing Group UK London}
}

@article{yoon2023,
  title={Handedness anomaly in a non-collinear antiferromagnet under spin--orbit torque},
  author={Yoon, Ju-Young and Zhang, Pengxiang and Chou, Chung-Tao and Takeuchi, Yutaro and Uchimura, Tomohiro and Hou, Justin T and Han, Jiahao and Kanai, Shun and Ohno, Hideo and Fukami, Shunsuke and others},
  journal={Nat. Mater.},
  volume={22},
  number={9},
  pages={1106--1113},
  year={2023},
  publisher={Nature Publishing Group UK London}
}

@article{miron2011,
  title={Perpendicular switching of a single ferromagnetic layer induced by in-plane current injection},
  author={Miron, Ioan Mihai and Garello, Kevin and Gaudin, Gilles and Zermatten, Pierre-Jean and Costache, Marius V and Auffret, St{\'e}phane and Bandiera, S{\'e}bastien and Rodmacq, Bernard and Schuhl, Alain and Gambardella, Pietro},
  journal={Nature},
  volume={476},
  number={7359},
  pages={189--193},
  year={2011},
  publisher={Nature Publishing Group UK London}
}

@Article{Velez2022,
author={V{\'e}lez, Sa{\"u}l
and Ruiz-G{\'o}mez, Sandra
and Schaab, Jakob
and Gradauskaite, Elzbieta
and W{\"o}rnle, Martin S.
and Welter, Pol
and Jacot, Benjamin J.
and Degen, Christian L.
and Trassin, Morgan
and Fiebig, Manfred
and Gambardella, Pietro},
title={Current-driven dynamics and ratchet effect of skyrmion bubbles in a ferrimagnetic insulator},
journal={Nat. Nanotechnol.},
year={2022},
month={Aug},
day={01},
volume={17},
number={8},
pages={834-841},
issn={1748-3395},
doi={10.1038/s41565-022-01144-x}
}
    
    \noindent\textbf{Methods}
\\
    {\color{blue}\noindent\textbf{Sample and device fabrication}\\
    The epitaxial thin-film BFO in this study was grown by pulsed laser deposition using a KrF excimer laser ($\lambda=248nm$) on an orthorhombic DyScO$_3$ (110) single-crystal substrate. For SIO/BFO heterostructure, a 40 nm BFO layer was grown at a substrate temperature of 800 $^\circ$C and an oxygen partial pressure of 0.15 mbar, followed by the growth of a 20 nm SIO layer at 650 $^\circ$C under oxygen partial pressure of 0.15 mbar. After the growth, the sample was cooled to room temperature in an oxygen environment of 300 mbar at a cooling rate of 20 $^\circ$C/min. For Pt/BFO devices, after the growth of BFO layer at identical conditions, we fabricatd the devices using photolithography. Afterwards, the Pt layer of 10 nm were deposited by magnetron sputtering.}  \\
\\
    %\noindent\textbf{Ferroelectric polarization switching}\\
    %The ferroelectric polarizations were switched using a ferroelectric tester (Precision Multiferroic, Radiant Technologies), with a fixed frequency of $10 kHz$ under room temperature. Two probes were placed on the Pt layer. \\
\\    
    {\color{blue}\noindent\textbf{Scanning NV magnetometry}\\
    Scanning-NV magnetometry was performed under ambient conditions with commercial all-diamond scanning probe tips containing single NV defects (SNVM, CIQTEK). The tip was integrated into a tuning-fork based atomic force microscope (AFM) combined with a confocal microscope optimized for single NV defect spectroscopy. Magnetic fields emanating from the sample are detected by recording the Zeeman shift of the NV defect's electronic spin sublevels through optical detection of the electron spin resonance. Experiments were performed with a NV-to-sample distance of approximately 50 nm and a bias magnetic field of 2 mT applied along the NV quantization axis which is at an angle of $\theta_{NV}\approx54.7\degree$ with respect to the surface normal.}  \\
\\
    {\color{blue}\noindent\textbf{Photocurrent measurement}\\
    Photocurrent measurement was conducted using a home-built confocal microscope (Extended Data Fig.~\ref{fig:Eig10}). A 405 nm solid-state laser (CNI Optoelectronics Technology Co., Ltd.) was used as the excitation source. The laser power was attenuated to 520 $\mu W$ using a density filter and the laser spot (diameter \textless 2 $\mu m$) was focused on the sample at normal incidence using a $\times 50$ Olympus objective (0.5 NA). The polarization state of the 405 nm laser was controlled by rotating a quarter-wave plate to switch between left-handed ($\varphi=45\degree$) and right-handed ($\varphi=135\degree$) circular polarization. Photocurrent mapping was carried out by scanning the laser spot across the sample using a two-axis galvanometer. A chopper (Sine Scientific Instruments Co., Ltd.) modulated the light at 133 Hz to facilitate lock-in detection. The photocurrent signal was amplified by an SR570 current preamplifier (Stanford Research Systems) and collected using an SR830 lock-in amplifier (Stanford Research Systems). The ferroelectric polarity of the BFO sample was switched by applying a poling voltage through a Keithley 2400 source meter (Tektronix, Inc.). All photocurrent measurements presented in the main text were performed at room temperature under ambient conditions.}\\    
\\
{\color{blue}{\noindent\textbf{Magnetic simulation}\\
\noindent The Heisenberg spin Hamiltonian of BFO has the form\upcite{Rahmedov2012,Huang2024}
\begin{align}\label{H_Heisenberg}
H=J\sum_{\langle ij\rangle}\mathbf{S}_i\cdot\mathbf{S}_j+\sum_{\langle ij\rangle}\mathbf{D}_1\cdot\mathbf{S}_i\times\mathbf{S}_j+\sum_{\langle ij\rangle^\prime}(\mathbf{D}_2\times\mathbf{e}_{ij})\cdot(\mathbf{S}_i\times\mathbf{S}_j)+K\sum_{i}(\mathbf{S}_i\cdot\mathbf{e})^2,
\end{align}
where $\mathbf{S}_{i,j}$ represents the unit spin vector at site $i$,$j$. The first term is the exchange interaction with strength $J$. The second and third terms are the homogeneous and inhomogeneous DMI with vectors $\mathbf{D}_1$ and $\mathbf{D}_2$, respectively, representing the DMI strength and direction. The fourth term denotes the uniaxial magnetic anisotropy with strength $K$ along the direction $\mathbf{e}$. The sum $\langle ij\rangle$ and $\langle ij\rangle^\prime$ runs over nearest and next-nearest neighboring, where $\mathbf{e}_{ij}$ is a unit vector connecting next-nearest neighboring sites. In BFO, the Fe atoms are arranged almost uniformly at the center of the unit cell with the lattice constant $a_0=0.4$ nm. The specific parameters for BFO are taken as $J=5.2$ meV, homogeneous DMI $D_1=0.05$ meV along $[110]$ direction, inhomogeneous DMI $D_2=0.3$ meV along $[111]$ direction, and anisotropy $K=-0.05$ meV along $[111]$ direction. In particular, we consider a cubic lattice of system size $L_x=100a_0$, $L_y=100a_0$, $L_z=2a_0$ with periodic boundary conditions along all three spatial directions. 

The magnetic ground state is obtained by numerically minimizing the Hamiltonian in Eq.~\ref{H_Heisenberg} using a custom Python implementation. This confirms a spin cycloid ground state propagating along $[\overline{1}10]$ direction, in agreement with the SNVM image in Fig.~\ref{fig:2}c. Moreover, the presence of homogeneous DMI ($D_1=0.05$meV) introduces a small canting between antiparallel neighboring spins, which hence results in a weak magnetization. On the other hand, we also minimize the Hamiltonian by using the steepest descent method from a designed magnetic state. The result affirms the existence of N\'eel type skyrmion in BFO  (Extended Data Fig.~\ref{fig:Eig3}).

The transition from spin cycloid to skyrmion can then be unveiled by solving the Landau-Lifshitz-Gilbert-Slonczewski equation\upcite{zhang2016}
{\color{blue}\begin{align}\label{LLGS}
\frac{d{\mathbf{S}_i}}{d{t}}=-\gamma\mathbf{S}_{i}\times\mathbf{H}_{eff}+\lambda\mathbf{S}_i\times\frac{d\mathbf{S}_i}{dt}+\xi\mathbf{S}_i\times\mathbf{p}\times\mathbf{S}_i,
\end{align}}
where $\mathbf{H}_{eff}=\partial{H}/\partial{\mathbf{S}_i}$ is the effective magnetic field, $\gamma$ is the gyromagnetic ratio, {\color{blue}$\lambda$} is the Gilbert damping parameter. The third term represents the Slonczewski spin-transfer torque arising from applied current with $\mathbf{p}$ the torque direction. The torque magnitude $\xi$ is proportional to the applied current density $J_c$ via the relation $\xi=\nu J_c$, where $\nu$ represents the spin-torque efficiency. To nucleate a skyrmion, we applied a 10ps current pulse to the spin cycloid ground state. Following the pulse, the system is allowed to relax sufficiently to a steady state using an enhanced damping constant of $\lambda=0.2$. In these simulations, the spin torque efficiency is taken as $\nu=10^3\rm{A}^{-1}$, and the magnitude of the inhomogeneous DMI ($D_2$) varies from $0.3$meV to $0.6$meV. To further determine the final magnetic state, we examine the spin distribution along with the N\'eel order topological charge defined as $Q=\sum_{i}\mathbf{L}_i\cdot\partial_x\mathbf{L}_i\times\partial_y\mathbf{L}_i/4\pi$, where $\mathbf{L}_i=(\mathbf{S}_i^A-\mathbf{S}_i^B)/2$ is the local N\'eel vector at site $i$. As shown in Fig.~\ref{fig:2}a, we identify three different regions on the $D_2-J_c$ plane agreeing with the experimental observations. The snapshots for the spin configuration during the nucleation under two different torque $J_c=1.0\times10^6\rm{A}\cdot\rm{cm}^{-2}$ and $J_c=1.1\times10^7\rm{A}\cdot\rm{cm}^{-2}$ are presented in Extended Data Fig.~\ref{fig:Eig11}.}}

\noindent{\textbf{Stray field simulation}}\\
\noindent The stray field in antiferromagnetic materials is generally vanishing because of the antiparallel magnetic moments. Fortunately, in BFO, the neighboring magnetic moments are slightly canted owning to $\mathbf{D}_1$, leading to a unique spatial-dependent stray field that can be detected by the NV center ({\color{blue}Extended Data Fig.~\ref{fig:Eig12}}). This stray field is mathematically connected to the canted magnetic moment through the relation\upcite{Dovzhenko2018,Tan2024}
\begin{align}\label{B_to_M}
\mathbf{B}(\mathbf{r},d)=-\frac{\mu_0M_s}{2}\begin{pmatrix}
-\alpha_z(d,t)*\partial_x^2,&-\alpha_z(d,t)*\partial_{y}\partial_x, &\alpha_{xy}(d,t)*\partial_x\\
-\alpha_z(d,t)*\partial_y\partial_x,&-\alpha_z(d,t)*\partial_{y}^2, &\alpha_{xy}(d,t)*\partial_y\\
\alpha_{xy}(d,t)*\partial_x,&\alpha_{xy}(d,t)*\partial_{y}, &\alpha_{z}(d,t)*\nabla^2
\end{pmatrix}\begin{pmatrix}m_x(\mathbf{r})\\m_y(\mathbf{r})\\m_z(\mathbf{r})\end{pmatrix},
\end{align}
where $t$ is the thickness of BFO, $d$ is the distance between the NV center and the BFO, $\mathbf{r}=(x, y)$ is the spatial coordinate. In the limit of $t\ll d$, the spatial resolution functions $\alpha_{xy}(d,t)=1/2\pi\cdot td/(d^2+r^2)^{3/2}$ and $\alpha_{z}(d,t)=1/2\pi\cdot t/(d^2+r^2)^{1/2}$. 
\\
The typical expression for a N\'eel-type antiferromagnetic skyrmion reads
\begin{align}\label{skx_expre}
\mathbf{L}(\mathbf{r})=\begin{pmatrix}\sin{\theta(\mathbf{r})}\cos{\phi(\mathbf{r})},&\sin{\theta(\mathbf{r})}\sin{\phi(\mathbf{r})},&\cos{\theta(\mathbf{r})}\end{pmatrix},
\end{align}
where $\phi(\mathbf{r})=\arctan{y/x}$, and $\theta(\mathbf{r})=\pi r/R$ when $r\le R$, while $\theta(\mathbf{r})=\pi$ when others with $R$ the skyrmion size. In the presence of $\mathbf{D}_1$\upcite{Albrecht2010}, the small net magnetization due to the canting of neighboring magnetic moments is $\mathbf{M}(\mathbf{r})=\mathbf{D}_1\times\mathbf{L}(\mathbf{r})$ ({\color{blue}Extended Data Fig.~\ref{fig:Eig12}}). In our simulations (Figs.~\ref{fig:2} and \ref{fig:3}), the skyrmion size $R=1.5$. The Antiferroelectricity in BiFeO3 Thin Films and the distance between NV center and the sample are taken as $d=2$, $t=0.05d$, respectively.
\\

\noindent{\textbf{BFO band calculation}}
\\
\noindent The first principles calculations are performed by using the Vienna ab initio simulation package (VASP) with the projector augmented wave (PAW) pseudopotential and the generalized gradient approximation (GGA) exchange-correlation potential. For self-consistency calculation, the plane-wave kinetic energy cutoff is set to be 600 eV and the total energy convergence criterion is less than $10^{-7}$ eV by sampling Brillouin zone with a $\mathbf{k}$-point mesh of $8\times 8\times 8$.

\noindent The lattice constant and atom positions of BFO have been fully relaxed until the Hellmann-Feynman forces on each atom are less than $10^{-3}$ eV/\AA, leading to a pseudo-cubic structure with $a=b=c=7.997$ \AA 
, and $\alpha=\beta=90.6^\circ$, $\gamma=89.3^\circ$. All the calculations are conducted in the absence of spin-orbit coupling, which results in collinear G-type antiferromagnetic order with zero magnetic moment. The ferroelectric polarization of BFO is evaluated to be 106 $\mu$C/cm$^2$ based on Berry phase method, consistent with previous first-principles calculation results.

\noindent\textbf{Data availability}\\
The data that support the findings of this study are available from the corresponding authors on reasonable request.\\
    
\noindent\textbf{Acknowledgements}\\
We are grateful for the fruitful discussions with Prof. Ramamoorthy Ramesh.\\    
    
\noindent\textbf{Author contributions}\\
X.C. supervised this study. G.W., H.B., Y.G. and K.S. carried out the synthesis of heterostructures and fabricated the devices. G.W., M.G., A.T. carried out the NV-imaging measurements. G.W., B.L., Y.G. carried out the CPGE measurements. Y.L., J.D., W.C., X.C., N. Z., J.Z. and X.X. performed the theoretical calculations. H.Z., J.Z., J.H., Y.W., H.J., X.Z., M.W., K.X., W.S., P.W., Q.L., C.S., Q.L. and M.L. gave suggestions on the experiments. All authors discussed the results and prepared the manuscript.\\
  
\noindent\textbf{Competing interests}\\
The authors declare no competing interests.\\

\noindent\textbf{Additional information}\\ Correspondence and requests for materials should be addressed to X.C.. 
Reprints and permissions information is available at http://www.nature.com/reprints.

    \setcounter{figure}{0}
    \captionsetup[figure]{labelfont={bf}, name={Extended Data Fig.}, labelsep=period}

\newpage
\begin{figure}[t!]
    	\centering
    	\includegraphics[width=1\textwidth]{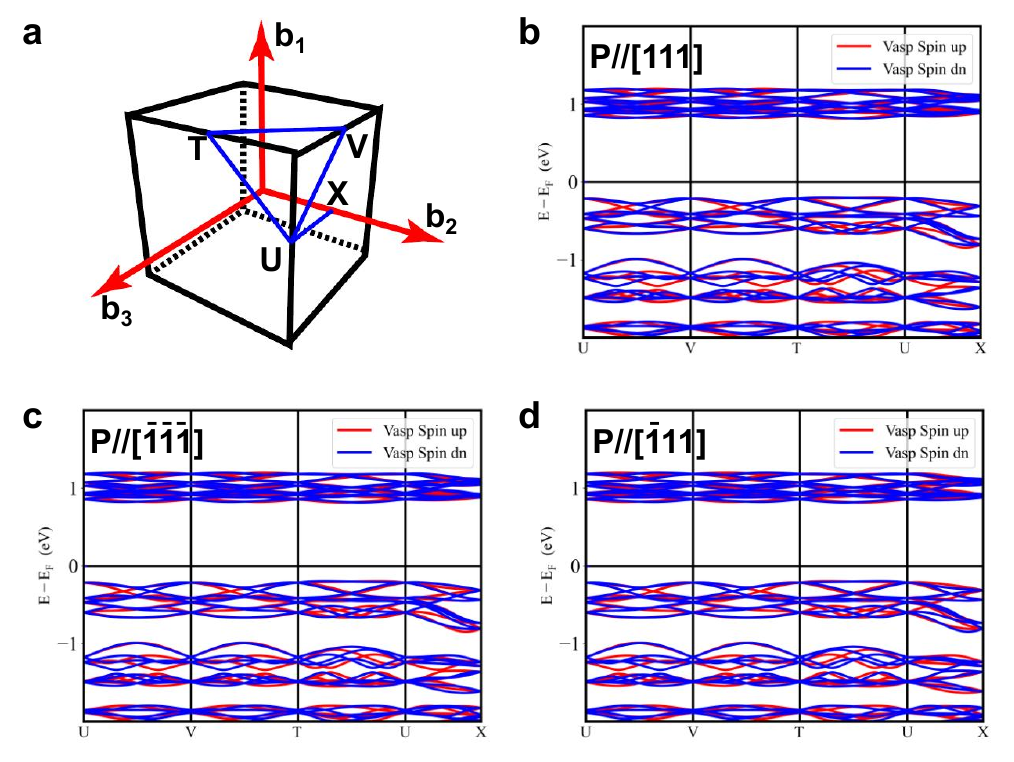}
        \caption{\textbf{BFO band structure. a,} First Brillouin zone of BFO. The high symmetric points are highlighted. \textbf{b-d,} First-principle calculated band structures for BFO with the ferroelectric polarization $\mathbf{P}$ pointing along [111] (\textbf{b}), [$\bar{1}\bar{1}\bar{1}$] (\textbf{c}) and [$\bar{1}11$]  (\textbf{d}) directions, respectively. 
        }
    	\label{fig:Eig1}
    \end{figure}

\begin{figure}[t!]
    	\centering
    	\includegraphics[width=1\textwidth]{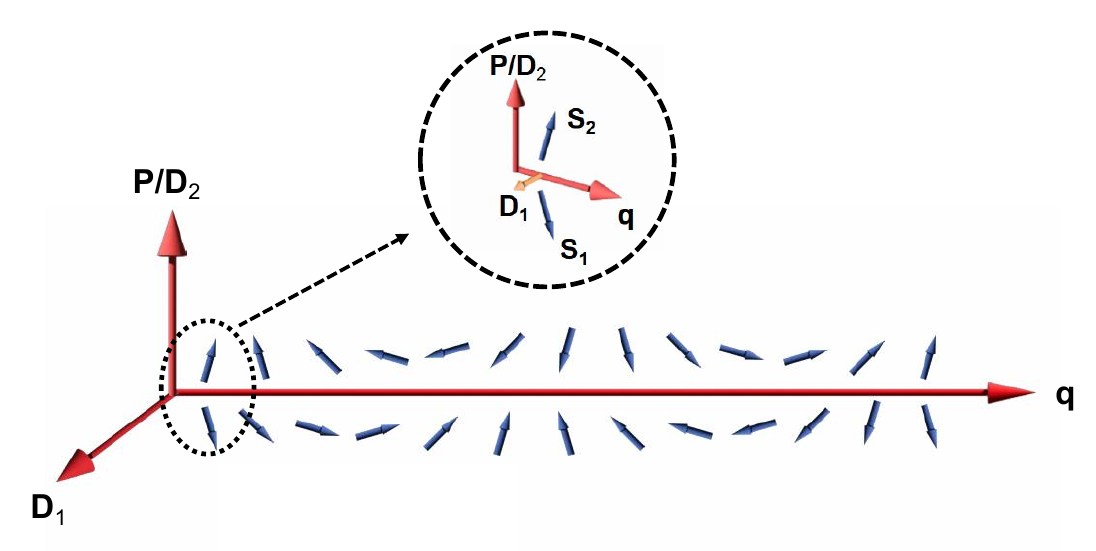}\caption{\textcolor{blue}{\textbf{The antiferromagnetic order of BFO.} A sketch of spin cycloid that propagates along $\mathbf{q}$= [$1\bar{1}0$]. Ferroelectric polarization $\mathbf{P}$ points along [111], while $\mathbf{D}_1$ and $\mathbf{D}_2$ represent homogeneous and inhomogeneous DMI, respectively, with {$\mathbf{D}_2\parallel\mathbf{P}$}, {$\mathbf{D}_2\perp\mathbf{q}$}, and {$\mathbf{D}_1\parallel\mathbf{S}_1\times\mathbf{S}_2$}.
        }}
    	\label{fig:Eig2}
    \end{figure}

\begin{figure}[t!]
    	\centering
    	\includegraphics[width=1\textwidth]{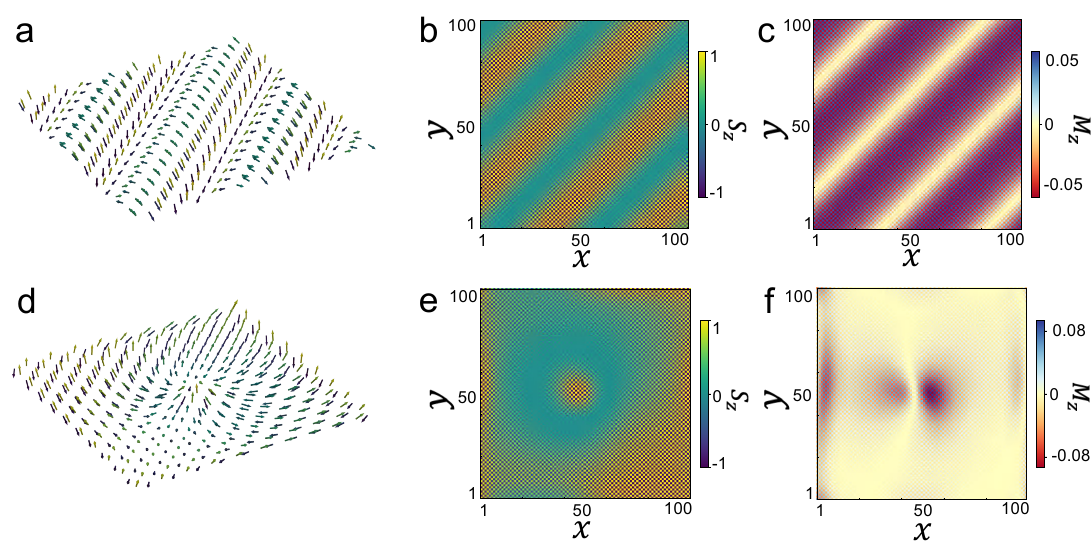}\caption{\textcolor{blue}{\textbf{Spin cycloid ground state and skyrmion state in BFO.} Three-dimensional plot (\textbf{a,d}), two-dimensional colormap (\textbf{b,e}) and the weak magnetization (\textbf{c,f}) between canted antiparallel spins on neighboring sites for the spin-cycloid ground state (\textbf{a-c}) and the N{\'e}el skyrmion state (\textbf{d-f}) in BFO. Here, the arrows in (\textbf{a,d}) represent the local spin direction while the color indicates $S_z$ value. Only 1 out of 25 ($5\times 5$) spin vectors is plotted in (\textbf{a,d}) for clarity.  
        }}
    	\label{fig:Eig3}
    \end{figure}
    
    \begin{figure}[t!]
    	\centering
    	\includegraphics[width=1\textwidth]{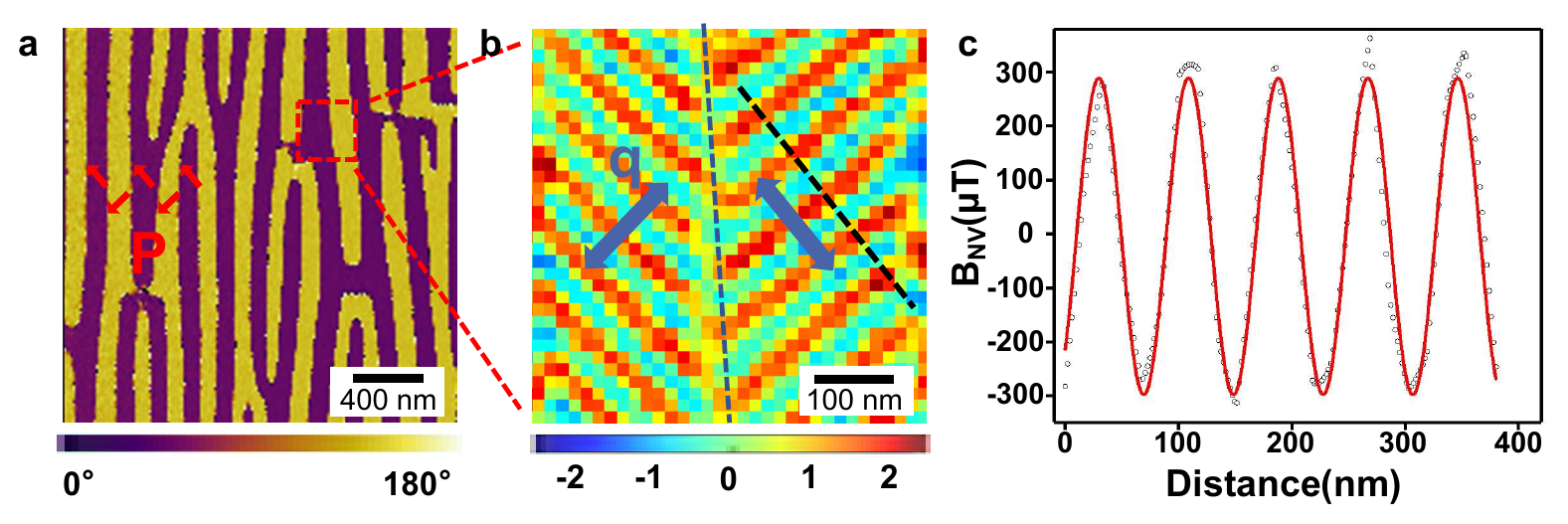}\caption{\textcolor{blue}{\textbf{Magnetoelectric coupling in BFO.} \textbf{a,} Lateral piezoresponse force microscopy (LPFM) image of a $\sim 100\text{nm}$ BFO film with 71° DWs grown on a (110) oriented DSO substrates. The red arrows indicate the in-plane projection of the ferroelectric polarization vector. \textbf{b,} Corresponding magnetic field distributions recorded from the dashed-red areas of the LPFM (\textbf{a,}) with the scanning-NV magnetometer (SNVM) operating in Full-B imaging mode. The blue arrow indicates the direction of $\mathbf{q}$ propagation. The scalebars in (\textbf{a}) and (\textbf{b}) are 400 nm and 100 nm, respectively. \textbf{a} Linecuts of the magnetic field distribution along the cycloid propagation direction (black dashed line in \textbf{b}).}
        }
    	\label{fig:Eig4}
    \end{figure}

\begin{figure}[t!]
    	\centering
    	\includegraphics[width=1\textwidth]{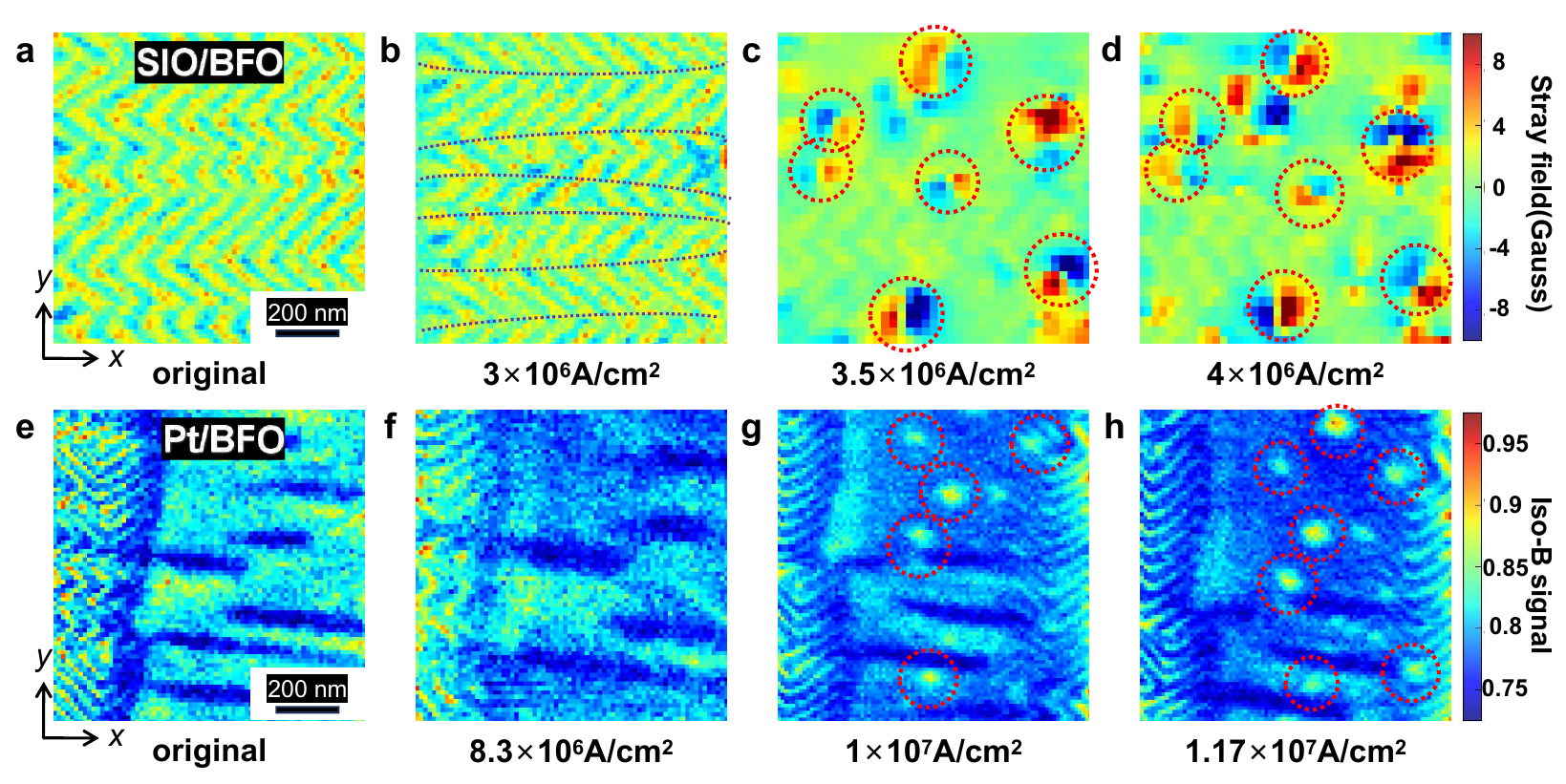}\caption{\textcolor{blue}{\textbf{Experimental study on pulse current regulation of BFO with different top electrodes a,} SNVM images of magnetic domains in SIO/BFO under applied pulsed current densities , showing the transformation of spin cycloids to skyrmions at $\sim 3.5\times 10^6$ A/cm$^2$. \textbf{e-h} The corresponding applied current density $J_s$ is at $\sim 1\times 10^7$ A/cm$^2$ for the top electrodes Pt/BFO. The scalebar in \textbf{a-h} is 200 nm. 
        }}
    	\label{fig:Eig5}
    \end{figure}

\begin{figure}[t!]
    	\centering
    	\includegraphics[width=1\textwidth]{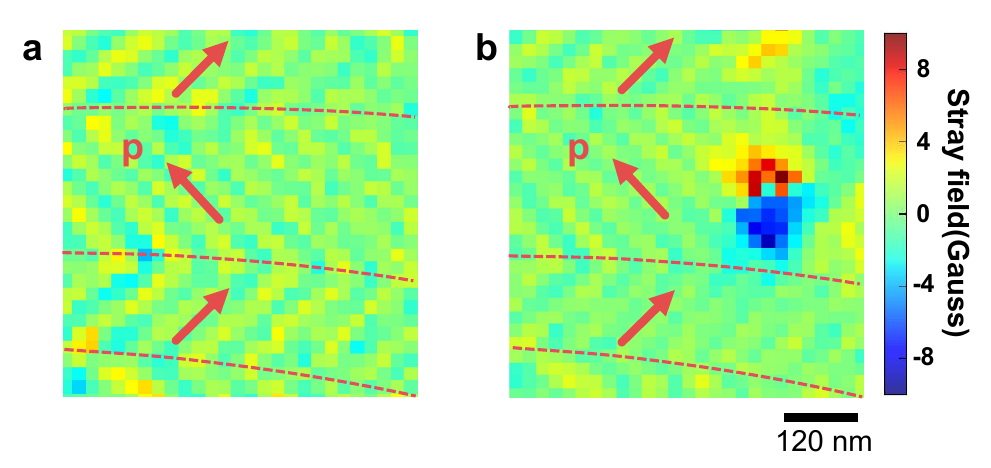}\caption{\textcolor{blue}{\textbf{Experimental study on electric polarization of BFO via SNVM.} \textbf{a, b,} SNVM images show the pulsed-current-driven transformation of magnetic domains in SIO/BFO. The red arrows indicate the in-plane projection of the ferroelectric polarization vector. The red dashed lines indicate the domain boundaries.
        }}
    	\label{fig:Eig6}
    \end{figure}

\begin{figure}[t!]
    	\centering
    	\includegraphics[width=1\textwidth]{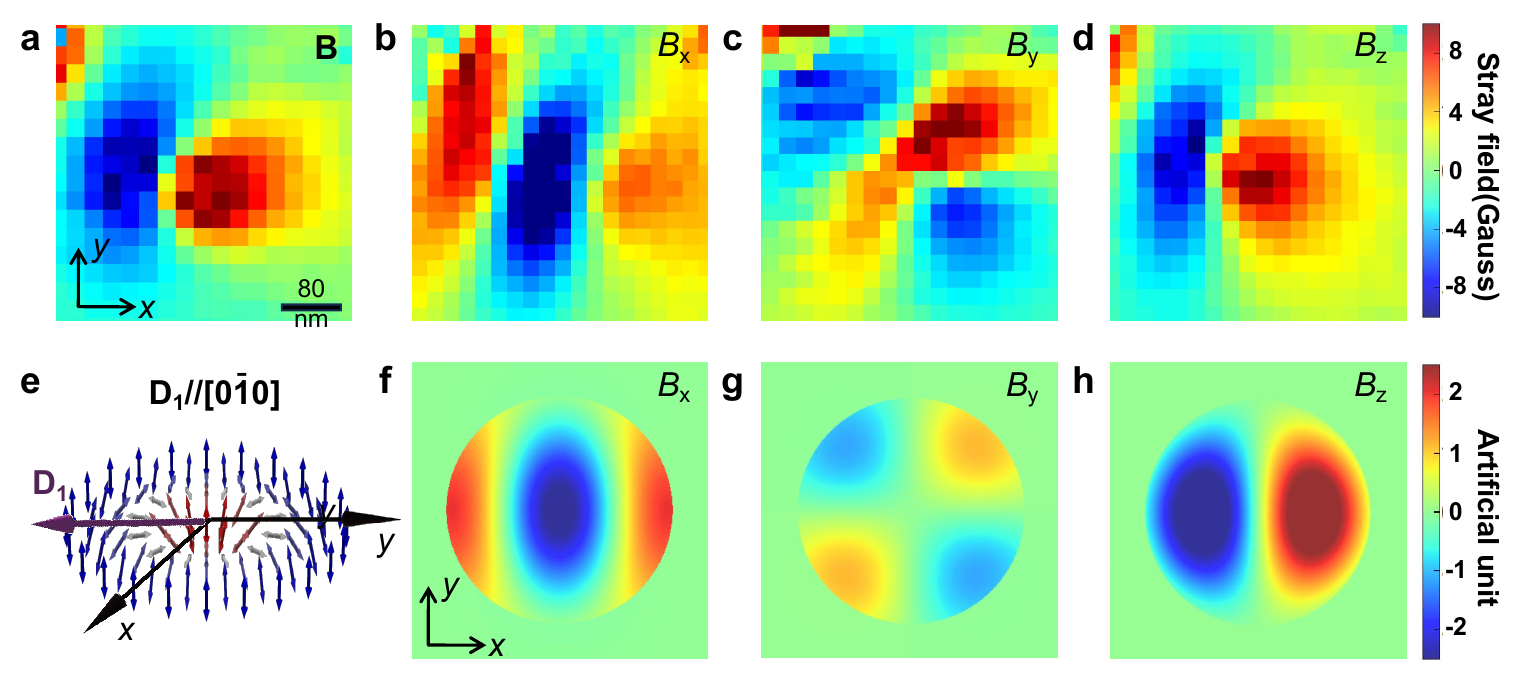}\caption{\textbf{Control of skyrmions by ferroelectric polarization} \textbf{a,} Various skyrmion phases, with DMI direction modulated by ferroelectric polarization. \textbf{b-d,} Stray field components reconstructed along the $x$, $y$, and $z$ directions based on panel \textbf{a}.  \textbf{e,} Simulated configuration of a skyrmion with DMI direction along [010].\textbf{ f-h,} Stray field distributions for the simulated skyrmion in \textbf{e} along the $x$, $y$, and $z$ directions, respectively.
        }
    	\label{fig:Eig7}
    \end{figure}

\begin{figure}[t!]
    	\centering
    	\includegraphics[width=1\textwidth]{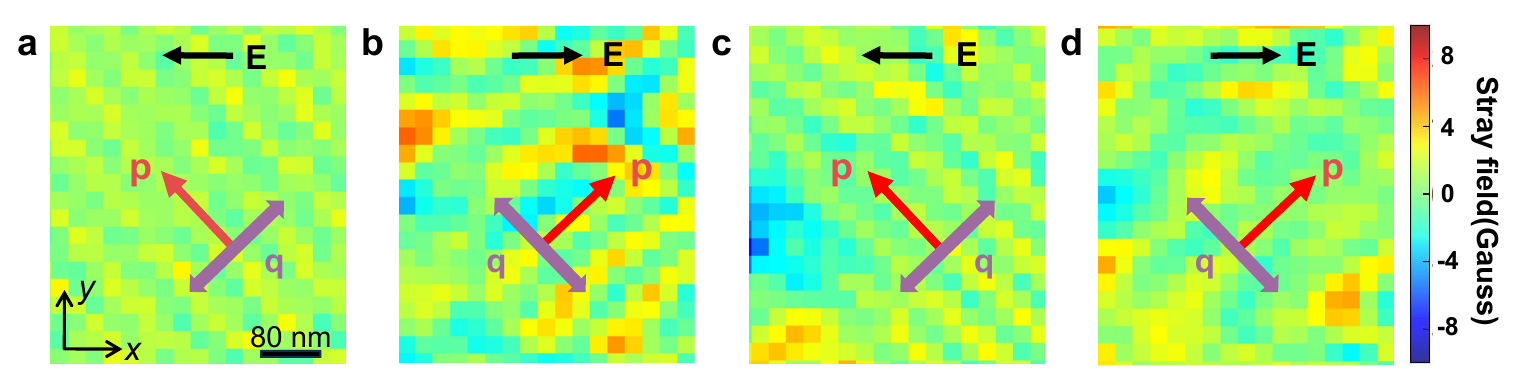}\caption{\textcolor{blue}{\textbf{Electrical control of the spin cycloid.} \textbf{a-d}  SNVM images of spin cycloid under E= -300 kV/cm (\textbf{a,c}) and E= 300 kV/cm (\textbf{b,d}),respectively. The polarization $\mathbf{P}$ align parallel to the directions marked by the red arrows in each panel, flipping by 90° under the electric field shown in (\textbf{a,c}) and (\textbf{b,d}). The purple arrows indicate the directions of $\mathbf{q}$ propagation, which are perpendicular to $\mathbf{P}$ directions. 
        }}
    	\label{fig:Eig8}
    \end{figure}

    \begin{figure}[t!]
            \centering
    	\includegraphics[width=1\textwidth]{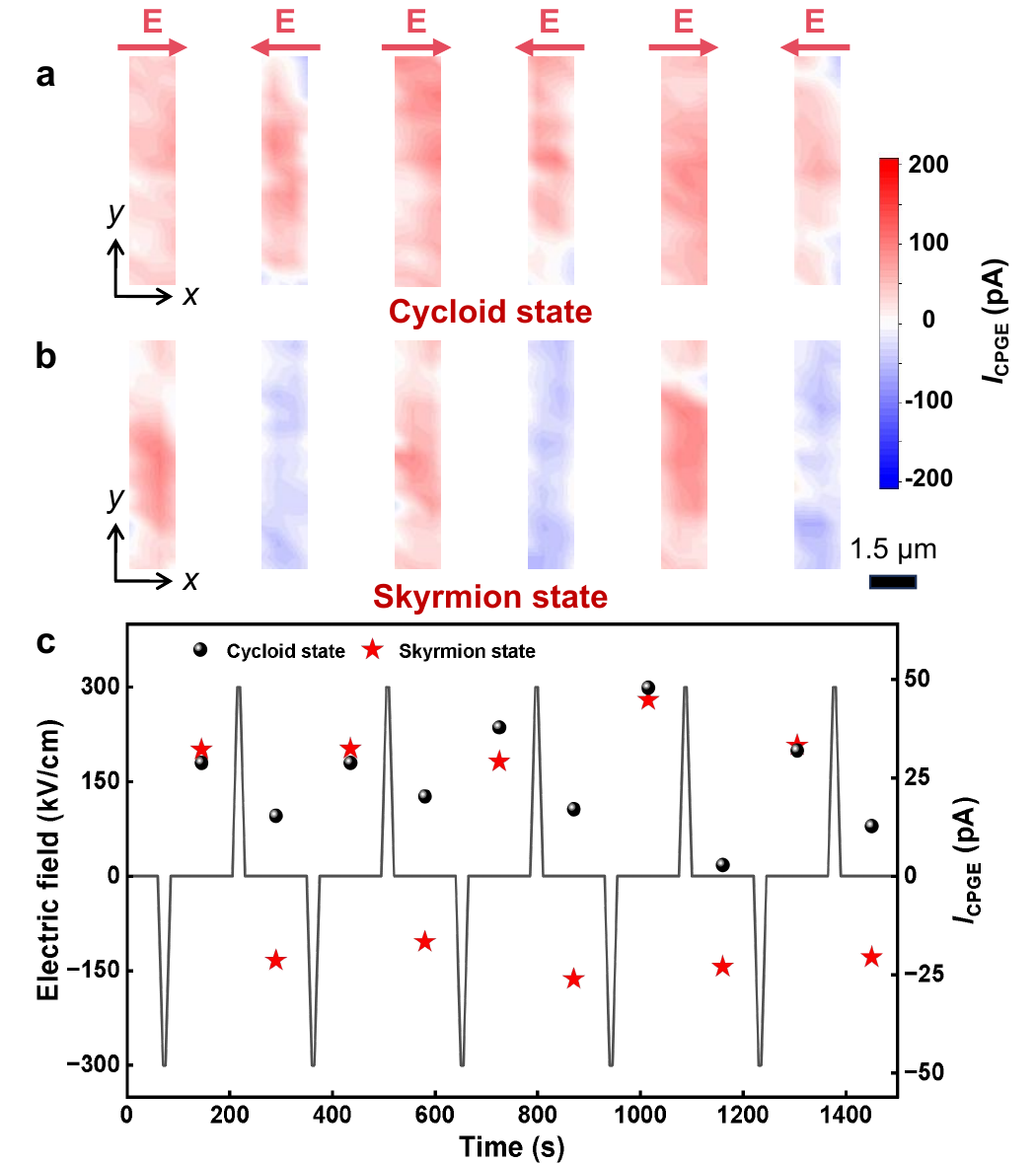}\caption{\textcolor{blue}{\textbf{Responses of the spin photocurrent ($I_{\mathrm{CPGE}}$) to electric-field modulation in cycloid and skyrmion states within the BFO system.} \textbf{a,b,} Mapping diagrams of the $I_{\mathrm{CPGE}}$ under electric-field modulation for the cycloid and skyrmion states, respectively. \textbf{c,} Black spheres and red asterisks denote the magnitudes of the overall average $I_{\mathrm{CPGE}}$ within the region of (\textbf{a}) and (\textbf{b}), respectively. For the cycloid state, the $I_{\mathrm{CPGE}}$ remains positive during the application of both positive and negative electric fields. Conversely, in the skyrmion state, the magnitude of the $I_{\mathrm{CPGE}}$ undergoes a regular transition in positive-negative polarity under electric-field modulation. The color bar indicates the magnitude of spin photocurrent in pA, and the scale bar is 1.5 $\mu m$.
        }}
    	\label{fig:Eig9}
    \end{figure}

\begin{figure}[t!]
            \centering
    	\includegraphics[width=1\textwidth]{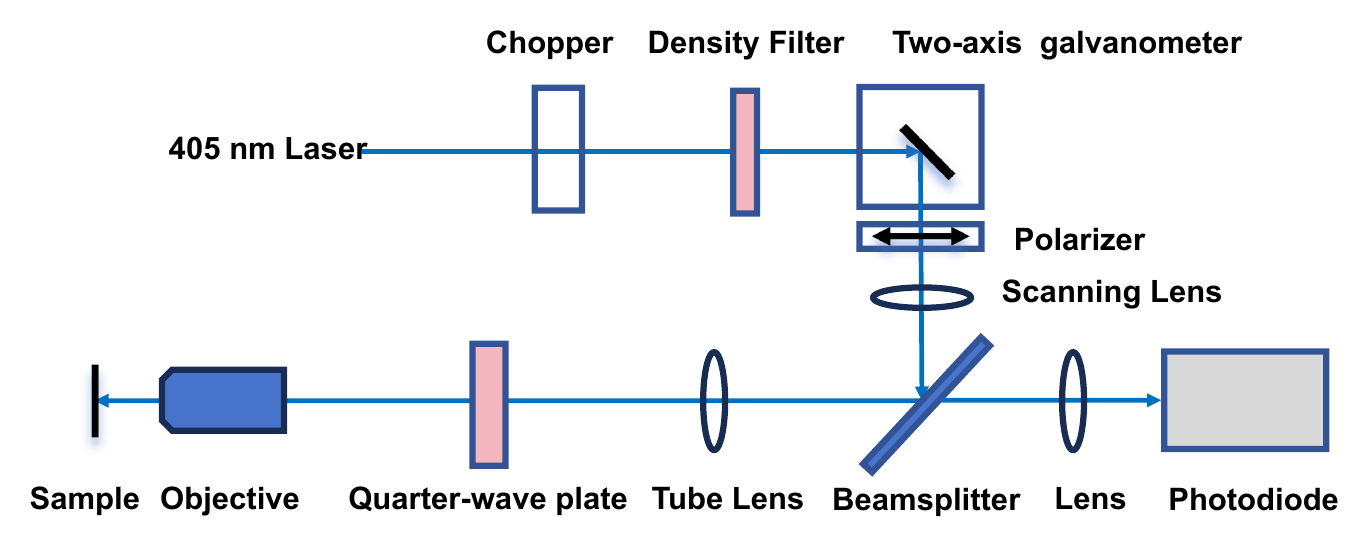}\caption{\textcolor{blue}{\textbf{Schematic figure for the optical measurement setup of the spin photocurrent
        measurement system.} A 405 nm laser spot was focused on the sample, with its polarization state tuned by a quarter-wave plate for circular polarization, and scanned using a two-axis galvanometer for photocurrent mapping.
        }}
    	\label{fig:Eig10}
    \end{figure}

\begin{figure}[t!]
    	\centering
    	\includegraphics[width=1\textwidth]{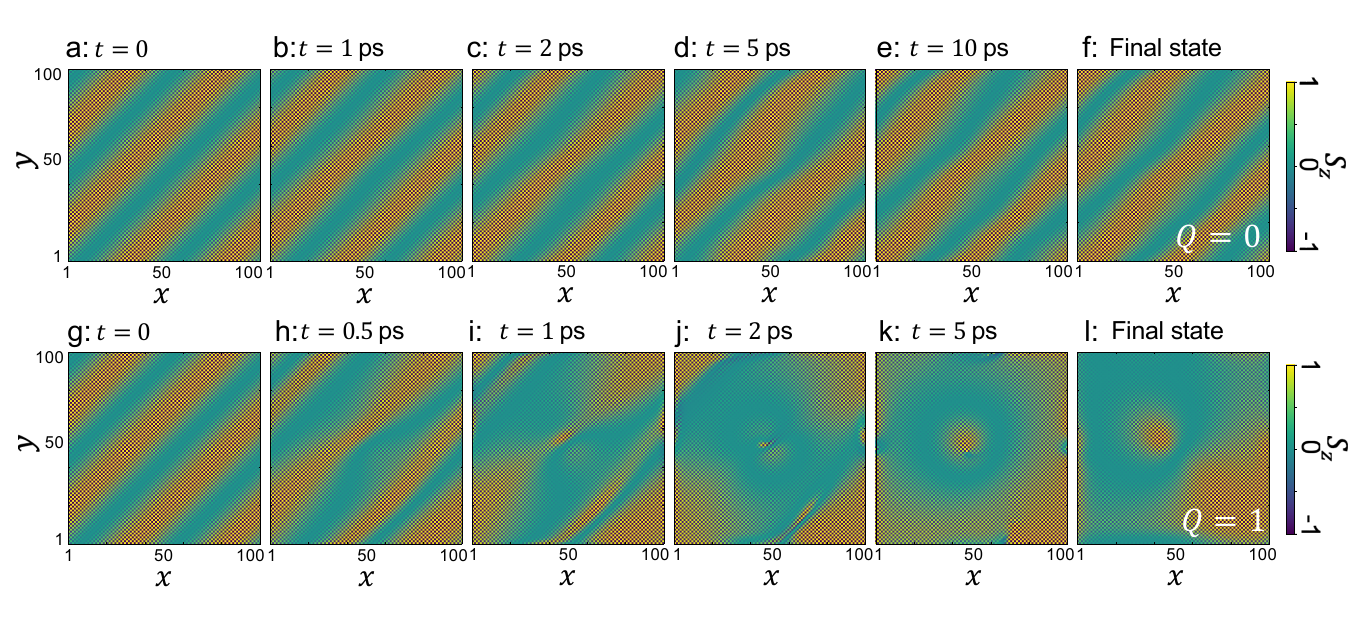}\caption{\textcolor{blue}{\textbf{Skyrmion nucleation in BFO via spin-orbit torque.} Time evolutions of the spin cycloid ground state at five representative time labeled ontop of each figure, and the final spin state after sufficient relaxation for BFO under a 10ps spin orbit torque of magnitude $J_c=1.0\times10^{6}\rm{A}\cdot\rm{cm}^{-2}$ (\textbf{a-f}) and $J_c=1.1\times10^{7}\rm{A}\cdot\rm{cm}^{-2}$ (\textbf{g-l}), respectively.  The Néel order topological charge for the final states in (\textbf{f}) and (\textbf{l}) are $Q=0$ and $Q=1$.
        }}
    	\label{fig:Eig11}
    \end{figure}

\begin{figure}[t!]
    	\centering
    	\includegraphics[width=1\textwidth]{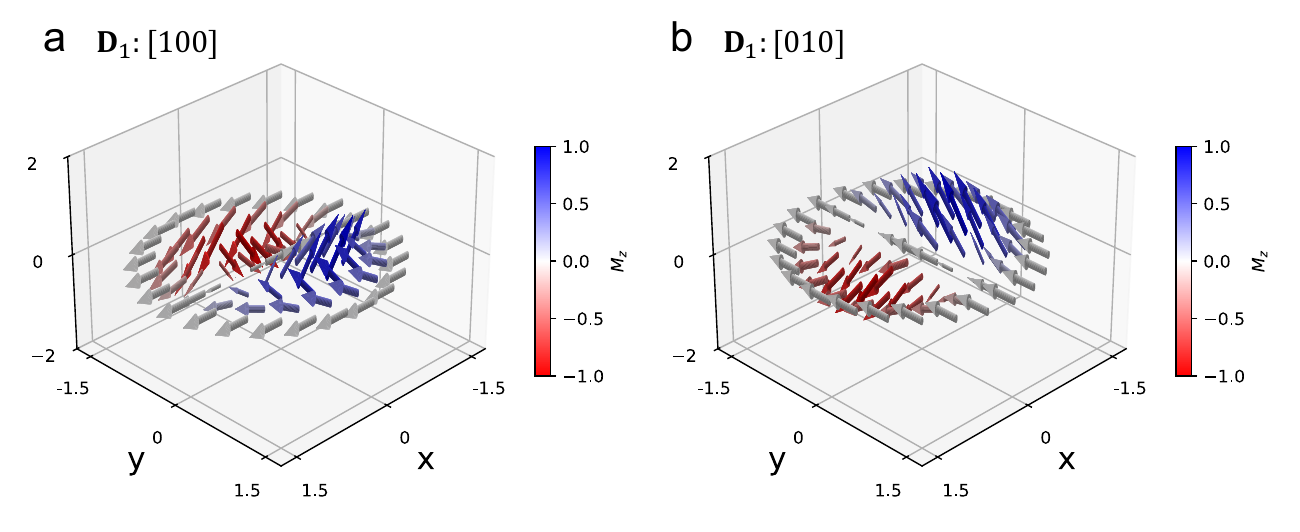}\caption{\textcolor{blue}{\textbf{ Canted moments for BFO skyrmion with different $\mathbf{D}_1$. a,} $\mathbf{D}_1$ aligns along [100]. {\textbf{b},} $\mathbf{D}_1$ aligns along [010].
        }}
    	\label{fig:Eig12}
    \end{figure}
    
\end{document}